\documentclass[oneside,twocolumn,aps,pre,preprintnumbers,bibnotes10pt,superscriptaddress,amsmath,amssymb]{revtex4-2}

\usepackage{graphicx}
\usepackage{subcaption}
\usepackage{float}
\usepackage[dvipsnames]{xcolor}
\usepackage{epstopdf}
\usepackage{amsmath}
\usepackage{braket}
\usepackage{amsthm}
\usepackage{amssymb}
\usepackage{amsfonts}
\usepackage{txfonts}
\usepackage{mathtools}
\usepackage{bm}
\usepackage{hyperref}
\usepackage{lipsum}
\usepackage[sort&compress]{natbib}
\usepackage{comment}
\usepackage{physics}
\usepackage{cancel}

\usepackage{lipsum}
\usepackage{booktabs}
\usepackage[normalem]{ulem}

\usepackage{tikz}
\usetikzlibrary{quantikz2,calc,arrows,chains,matrix,positioning,scopes}

\usepackage{letltxmacro}
\LetLtxMacro{\originaleqref}{\eqref}
\renewcommand{\eqref}{Eq.~\originaleqref}

\newcommand{\probP}{\text{I\kern-0.15em P}}

\usepackage{tikz}
\makeatletter
\newcommand\mathcircled[1]{%
  \mathpalette\@mathcircled{#1}%
}
\newcommand\@mathcircled[2]{%
  \tikz[baseline=(math.base)] \node[draw,circle,inner sep=0.5pt] (math) {$\m@th#1#2$};%
}
\makeatother

\begin{document}

\title{Macroscopic Fokker-Planck equation from microscopic Glauber dynamics for the Nagle-Kardar model}

\author{Jean-Fran\ifmmode \mbox{\c{c}}\else \c{c}\fi{}ois de Kemmeter}
\affiliation{Istituto Nazionale di Ottica (INO), Consiglio Nazionale delle Ricerche (CNR), Largo Enrico Fermi 6, I-50125 Firenze, Italy.}
\affiliation{Department of Mathematics, Royal Military Academy, Brussels, Belgium.}

\author{Stefano Ruffo}
\affiliation{Istituto dei Sistemi Complessi (ISC), Consiglio Nazionale delle Ricerche (CNR), Via Madonna del Piano 10, I-50019 Sesto Fiorentino, Italy.}
\affiliation{INFN, Sezione di Firenze, I-50019 Sesto Fiorentino, Italy.}
\affiliation{SISSA, via Bonomea 265, I-34136 Trieste, Italy.}

\author{Stefano Gherardini}
\affiliation{Istituto Nazionale di Ottica (INO), Consiglio Nazionale delle Ricerche (CNR), Largo Enrico Fermi 6, I-50125 Firenze, Italy.}

\begin{abstract}
In the companion Letter~\cite{deKemmeterLetter2025} we have highlighted the dynamical universality class of the Nagle-Kardar model, where a mean-field interaction is added to the one-dimensional nearest-neighbor Ising model. Starting from the microscopic Glauber dynamics, this paper provides a complete derivation of the Fokker-Planck equation that describes the time evolution of the macroscopic variables---magnetization and defect density---appearing in the model's Hamiltonian.
The study of the Langevin equation, associated with the Fokker-Planck equation, allowed us to prove that the model belongs to the universal class of systems with diffusive dynamics and non-conserved order parameter (model A)~\cite{deKemmeterLetter2025}. To this goal, we used several features of the model at equilibrium, including the phase diagram and the fluctuations of macroscopic variables, which are here discussed for completeness. The derivation of the Fokker-Planck equation requires the solution of some combinatorial problems that appear in the counting of configurations at an intermediate level between the microscopic Glauber dynamics and the macroscopic Fokker-Planck one.
Finally, we apply both the Glauber and the Langevin dynamics to the study of the average first passage time between local equilibrium states. We confirm that this time obeys an exponential Arrhenius law in terms of system's size, offering a direct link between microscopic energy landscapes and macroscopic relaxation mechanisms.
\end{abstract}

\maketitle

\section{Introduction}

Interest in the study of systems with long-range interactions has grown significantly in recent years~\cite{campa2009statistical,defenu2023long}. The case of a slowly-decaying interaction, where the potential $V(r)$ behaves as $V(r) \sim r^{-\alpha}$ at large distance $r$, poses many technical difficulties both at the level of analytical  approaches~\cite{Dyson1969,ThoulessPR1969,Dyson1971} and of numerical simulations~\cite{LuijtenPRB1997,Tomita2009, muller2025efficient}.

Moreover, in several physical systems one observes a coexistence and competition of short- and long-range interactions. In some instances, these cases can be treated using simpler approaches that nevertheless allow for the calculation of various physical quantities, such as averages of macroscopic quantities, fluctuations, and correlations~\cite{campa2014physics}. A notable example is the so-called Nagle-Kardar model~\cite{nagle1970ising,kardar1983crossover}, where a term with mean-field interaction is added to the Hamiltonian of the short-range Ising model. The companion Letter of this paper illustrates in detail the finite-time and finite-size behaviors of this model in the critical region~\cite{deKemmeterLetter2025}.

In this paper, we begin by recapitulating the solution of the one-dimensional Nagle-Kardar model in the canonical ensemble discussing in detail the phase diagram (Section~\ref{sec2}). The latter shows the coexistence of first- and second-order phase transition lines, as well as of a tricritical point. 
We omit a discussion of ensemble inequivalence, which would otherwise open a different chapter of interesting phenomena~\cite{mukamel2005breaking,CampaJPAMT2025}.
We go beyond the analysis of Nagle and Kardar obtaining: {\it (i)} the exact probability distribution of total magnetization and number of defects for a finite number of spins (Section~\ref{sec3}), {\it (ii)} the large-deviation behavior of macroscopic variables, i.e., magnetization per spin and defect density (Section~\ref{sec4}), {\it (iii)} an explicit formula for the spin correlation function (Section~\ref{sec5}). When discussing finite-size scaling in the companion Letter~\cite{deKemmeterLetter2025}, we have used the knowledge of the static exponents $\beta$ and $\nu$. The exponent $\beta$ is derived in Section~\ref{sec6}, where we also justify the value of the exponent $\nu$ by using both finite-size scaling~\cite{FisherPRL1972,Barber1983,Privman1990Editor} and Landau-Ginzburg theory~\cite{Goldenfeld1992Lectures}.

However, the main result of this paper is the derivation of the time evolution of magnetization and defect density at the macroscopic level from the microscopic Glauber dynamics~\cite{glauber1963time}. We begin by introducing the master equation in Section~\ref{sec7}. Then, we derive the coarse-grained transition rates under the assumption that the system dynamics explores the microstates sufficiently fast compared to the relaxation time of the macroscopic variables. For large system sizes, we deduce the Fokker-Planck equation for the intensive variables, i.e., magnetization and defect densities (Section~\ref{sec8}). In order to characterize the time evolution of these variables, we also derive the corresponding Langevin equation (Section~\ref{sec8}). This derivation turns out to be crucial when discussing a further important feature of the Nagle-Kardar model, namely, the exponential divergence in the number of spins of the average first-passage time among minima of the effective large-deviation potential, which is the subject of Section~\ref{sec9}~\cite{griffiths1966relaxation,antoni2004first,mukamel2005breaking,chavanis2005lifetime,chavanis2026thermal}.

\section{Free energy and phase diagram in the canonical ensemble}\label{sec2}

Let us consider the Hamiltonian of the Nagle-Kardar model:
\begin{equation}\label{eq:Hamiltonian}
    H = -J \sum_{i=1}^N s_i s_{i+1} - \frac{K}{2N} \left(\sum_{i=1}^N s_i\right)^2 - h \sum_{i=1}^N s_i,
\end{equation}
where $N$ is the number of spins, $s_i \in\{-1,+1\}$, and the parameters $J, K$ and $h$ represent the nearest-neighbor coupling strength, the mean-field coupling strength and the external field, respectively. Throughout the paper, periodic boundary conditions are assumed, i.e., $s_{N+1} = s_1$.

The Hamiltonian (\ref{eq:Hamiltonian}) can be equivalently expressed in terms of both the total magnetization $M=\sum_{i=1}^N s_i$ and the number of defects. A defect is a formal representation of the presence of a pair of adjacent spins with opposite signs. Under periodic boundary conditions, the number of defects is always even and given by $2S = \sum_{i=1}^N (1-s_i s_{i+1})/2$. In terms of $M$ and $S$, the Hamiltonian \eqref{eq:Hamiltonian} becomes:
\begin{equation}
    H = -J(N-4S) - \frac{K}{2N} M^2 - h M.
\end{equation}
The total magnetization $M$ ranges from $-N$ to $+N$ in steps of $2$, while the number of defects $2S$ takes values from $2(1-\delta_{\vert M\vert,N})$ to $N-\vert M\vert $ also in steps of $2$ ($\delta_{ij}$ is the Kronecker delta). Given a configuration with magnetization $0\leq  M  < N$, the minimum number of defects is $2$ that occurs when all down-spins are contiguous. On the other hand, the number of defects is maximized when each down-spin is isolated, i.e., surrounded by up-spins on both sides, which leads to $N-M$ defects (since there are $(N-M)/2$ down-spins). A similar argument holds for negative magnetization $-N<M<0$. In the fully polarized cases $M=\pm N$, $2S=0$ as no defect is present.

The equilibrium probability to observe a spin configuration $\mathcal{C}=(s_1,\cdots,s_N)$ is given by the Gibbs measure
\begin{equation}\label{eq:Gibbs_weight}
    \omega_N(\mathcal{C}) = \frac{e^{-\frac{H(\mathcal{C})}{T}}}{Z_N} = \frac{e^{\frac{J}{T}(N-4S)+\frac{K}{2NT}M^2 + \frac{h}{T}M}}{Z_N},
\end{equation}
where $Z_N=\sum_{\mathcal{C}} \omega_N(\mathcal{C})$ denotes the partition function, and $T$ is temperature with the Boltzmann constant $k_B$ set to $1$. The latter can be evaluated in the thermodynamic limit $N \rightarrow \infty$ using the Hubbard-Stratonovich transformation. Specifically, using the identity
\begin{equation}\label{eq:int}
    e^{\frac{K}{2NT} \big(\sum_{i=1}^N s_i\big)^2} = \sqrt{\frac{N K}{2\pi T}}
    \int_{-\infty}^{+\infty}dx\, e^{- \frac{N K}{2T} x^2} e^{ \frac{K x}{T} \sum_{i=1}^N s_i},        
\end{equation}
the partition function reads as
\begin{equation}
    \begin{split}
    Z_N &= \sum_{\mathcal{C}} \sqrt{\frac{N K}{2\pi T}}
		\int_{-\infty}^{+\infty}dx\, e^{\frac{J}{T}\sum_{i=1}^N s_i s_{i+1}} e^{- \frac{N K}{2T} x^2} e^{\frac{1}{T}(K x+h) \sum_{i=1}^N s_i}\\
    &= \sqrt{\frac{N K}{2\pi T}}
		\int_{-\infty}^{+\infty}dx\, e^{- \frac{N K}{2T} x^2}\, \text{Tr}\left( V^N \right),
    \end{split}
\end{equation}
where the $2\times 2$ transfer matrix $V$ is given by
\begin{equation}
    V = \begin{pmatrix}
        e^{(J + Kx+h)/T} & e^{-J/T} \\
        e^{-J/T} & e^{(J - Kx - h)/T} 
    \end{pmatrix}.
    \label{eq:Trmat}
\end{equation}
The eigenvalues $\lambda_{\pm}(x)$ of the transfer matrix $V$ are:
\begin{eqnarray}\label{eq:lambda}
    \lambda_\pm(x) &=& e^{J/T} \cosh\left(\frac{1}{T}(Kx+h)\right) \nonumber \\
    &\pm& \sqrt{e^{2J/T}\sinh^2\left(\frac{1}{T}(Kx+h)\right) + e^{-2J/T}}.
\end{eqnarray}
In the large-$N$ limit, the eigenvalue $\lambda_{+}(x)$ is dominant, with the result that
\begin{equation}
    Z_N \sim \sqrt{\frac{N K}{2\pi T}} e^{-\frac{N}{T} \left[\frac{K}{2} {x^*}^2 - T\log\left(\lambda_+(x^*)\right)\right]},
\end{equation}
where $x^*$ is the absolute minimum of the free-energy density $\mathcal{F}(T, J,  K, h;x)$ given by
\begin{equation}
    \mathcal{F}(T, J, K, h;x) = \frac{K}{2} {x}^2 - T\log\left( \lambda_+(x) \right).
\end{equation}
\begin{figure}[t!]
    \centering
    \includegraphics[width=0.98\linewidth]{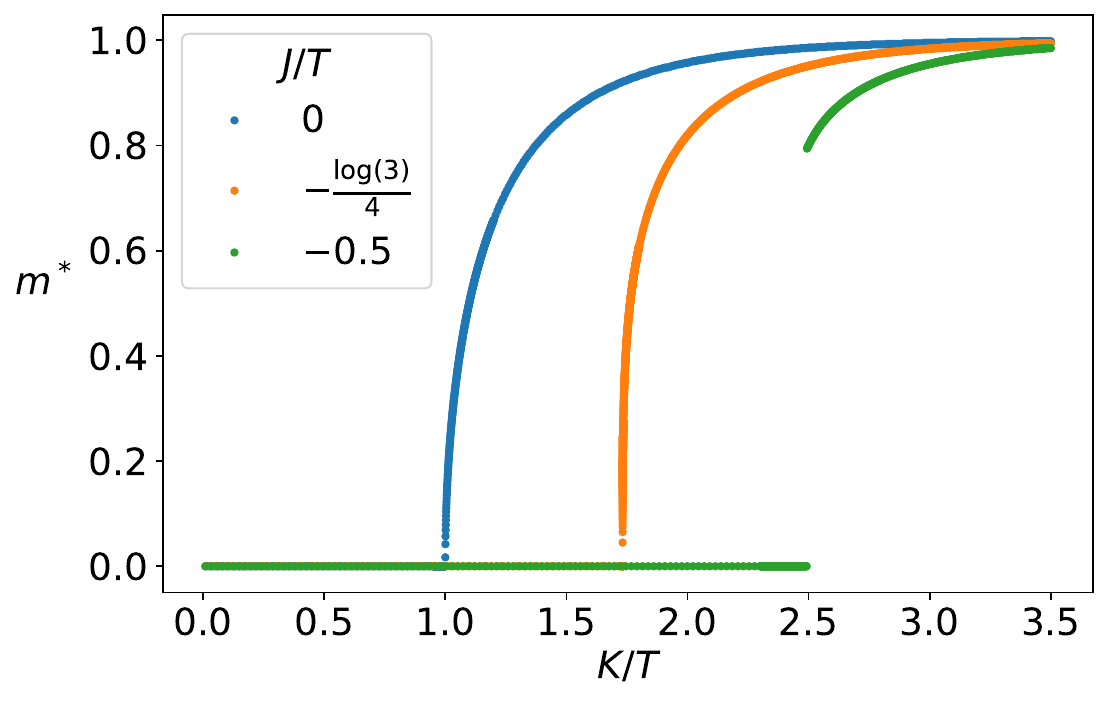}
    \caption{
    Equilibrium magnetization $m^*$ as a function of $K/T$, for $J/T = 0$ (second-order transition), $J/T = -\log(3)/4$ (tricritical point) and $J/T = -0.5$ (first-order transition).
    }
    \label{SMfig:Phase_diagram}
\end{figure}
Explicit computations show that $x^*$ is solution of the implicit equation 
\begin{equation}
    x^* = \frac{\sinh\Big( (h+Kx^*)/T \Big)}{\sqrt{\sinh^2 \Big( (h+Kx^*)/T \Big) + e^{-4 J/T}} }.
    \label{SMeq:implm}
\end{equation}
The free energy per site is then given by $f(T, J, K, h) = \mathcal{F}(T, J, K, h;x^*)$. The absolute minimum $x^*$ coincides with the magnetization per site
\begin{equation}
    m^* := \lim_{N\rightarrow +\infty}\frac{1}{N} \sum_{i=1}^N \langle s_i \rangle.
\end{equation}
Indeed, we have:
\begin{eqnarray}
    m^{*} &=& -\frac{\partial }{\partial h} f(T, J, K, h) = -\frac{\partial }{\partial h} \mathcal{F}(T, J, K, h;x^*) \nonumber\\ 
    &=& \frac{T}{ K} \frac{\partial }{\partial x}\log\left( \lambda_{+}(x)\right)\vert_{x=x^*} = x^*.
\end{eqnarray}
The magnetization $m^*$ is thus the global minimum of the free-energy density $\mathcal{F}(T, J, K, h;m)$, whose expression is:
\begin{eqnarray}
    \mathcal{F}(T, J, K, h;m) &=& \frac{K}{2} m^2 - T\log\left(\lambda_+(m)\right) \nonumber \\
    &=& \frac{K}{2}m^2 - J - T\log
    \Big[ \cosh\Big( (Km+h)/T \Big)\nonumber \\
    &&+ \sqrt{\sinh^2\Big( (Km+h)/T \Big) + e^{-4 J/T}}
    \Big].
\end{eqnarray}
Formally, \eqref{SMeq:implm} may admit several solutions. Among them, only the one that minimizes $\mathcal{F}(T, J, K, h;m)$ corresponds to the equilibrium magnetization in the thermodynamic limit. In the absence of the external field ($h=0$), $\mathcal{F}(T, J, K, h=0;m)$ is symmetric in $m$ and its Taylor expansion around $m=0$ reads as
\begin{eqnarray}
    &&\mathcal{F}(T, J, K, h=0;m) = - T\log\big(2\cosh(J/T)\big) +\nonumber \\
    &&+\frac{1}{2} K \Big(1-\frac{K/T}{e^{-2 J/T}}\Big) m^2 +\frac{K^4 (-1+3e^{4 J/T})}{24 \, T^{3} e^{-2 J/T}}m^4 + O(m^6).\quad
\end{eqnarray}

A second-order phase transition occurs when the coefficient of the quadratic term vanishes, i.e., when $K/T = e^{-2 J/T}$~\cite{nagle1970ising, kardar1983crossover}. This second-order line ends at a \textit{tricritical point}, where both the quadratic and quartic coefficients of $\mathcal{F}(T, J, K, h;m)$ in $m$ simultaneously vanish. This condition corresponds to set
\begin{equation}
    \Big( J/T, K/T \Big) = \left( -\frac{\log(3)}{4}, \sqrt{3} \right).
\end{equation}
For values $J/T < -\frac{\log(3)}{4}$ and $K/T > \sqrt{3}$, the system exhibits a line of first-order phase transition that is characterized by a discontinuous jump in the magnetization, as depicted by the green dots in Fig.~\ref{SMfig:Phase_diagram} where the different behaviors of $m^*$ as a function of $K$ ($T=1$) are plotted.

\section{Distribution of magnetization and number of defects for finite size $N$}\label{sec3}

In this section we present an exact expression for the equilibrium joint probability distribution $P_N(M,2S)$ of configurations of $N$ spins with total magnetization $M$ and $2S$ defects.

Given the statistical weight $\omega_N(M,2S)$ [\eqref{eq:Gibbs_weight}] of any configuration with total magnetization $M$ and $2S$ defects at equilibrium, the joint probability $P_N(M,2S)$ of observing a configuration with magnetization $M$ and $2S$ defects reads as
\begin{eqnarray}
    P_N(M,2S) &=& \Omega_N(M,2S)\omega_N(M,2S)\nonumber \\
    &=& \Omega_N(M,2S)\frac{e^{\frac{J}{T}(N-4S)+\frac{K}{2NT}M^2 + \frac{h}{T}M}}{Z_N},
\end{eqnarray}
where $\Omega_N(M,2S)$ enumerates the distinct configurations with magnetization $M$ and $2S$ defects. The expression of $\Omega_N(M,2S)$ can be computed explicitly~\cite{antal2004probability}, and it is given by
\begin{equation}
    \Omega_N(M,2S) = \binom{\frac{N+M}{2}-1}{S-1} \binom{\frac{N-M}{2}}{S} + \binom{\frac{N-M}{2}-1}{S-1} \binom{\frac{N+M}{2}}{S} + \delta_{M,\pm N}\delta_{S,0},
    \label{eq:bincoeff}
\end{equation}
where the binomial coefficients are understood to be zero whenever any of their arguments are negative or non-integer.

The first two terms on the right-hand-side of \eqref{eq:bincoeff} are obtained as follows. In a configuration of $N$ spins with magnetization $M$, there are $\frac{N+M}{2}$ up-spins and $\frac{N-M}{2}$ down-spins. In the presence of $2S$ defects, exactly $S$ of them lie to the right (left) of a down-spin, while the other lie to the left (right) of an up-spin. Suppose the first spin of the configuration is a spin $-1$, as in the following example with $N=11$ [plus a spin, encircled, determined by the boundary condition (b.c.)], $M=-3$ and $2S=6$:
\begin{eqnarray*}
    &&\underbrace{ 
    \left\{
        \overset{\upsilon_1}{-} \,,\, \overset{\upsilon_2}{---} \,,\, \overset{\upsilon_3}{--} \,,\, \overset{\upsilon_1}{-} 
    \right\}
    }_{\frac{N-M}{2} \text{ down-spins}}
    \quad\text{and}\quad
    \underbrace{
    \left\{
        \overset{\mu_1}{++} \,,\, \overset{\mu_2}{+} \,,\, \overset{\mu_3}{+} 
    \right\}
    }_{\frac{N+M}{2} \text{ up-spins}}\nonumber \\
    &&\longrightarrow \quad
    \underbrace{
    \overset{\upsilon_1}{\mathcircled{-}}\vert\overset{\mu_1}{++}\vert\overset{\upsilon_2}{---}\vert\overset{\mu_2}{+}\vert\overset{\upsilon_3}{--}\vert\overset{\mu_3}{+}\vert \overset{\upsilon_1}{-\mathcircled{-}}
    }_{N \text{ spins (periodic b.c.)}}
\end{eqnarray*}
where the vertical bar $\vert$ stands for a defect.
There are $\binom{\frac{N-M}{2}}{S}$ distinct ways of placing $S$ defects among the $\frac{N-M}{2}$ down-spins, partitioning the down-spins into $S$ consecutive groups of sizes $\upsilon_1,\upsilon_2,\ldots,\upsilon_S$, each of them containing at least one spin. In the example, $\upsilon_1=2,\upsilon_2=3,\upsilon_3=2$. Similarly, one has to distribute the $S$ defects among the $\frac{N+M}{2}$ up-spins, with the constraint that the last up-spin is immediately followed by a defect. This can be done in $\binom{\frac{N+M}{2}-1}{S-1}$ distinct ways. In the example, $\mu_1=2, \mu_2=1$ and $\mu_3=1$. The resulting configuration is formed by alternating the groups of sizes $\upsilon_i$ and $\mu_i$ for $i=1,\ldots,S$. The total number of such configurations corresponds to the first term of \eqref{eq:bincoeff}. The second  term in \eqref{eq:bincoeff} is derived similarly, with the assumption that the first spin is an up-spin.

The last term of the right-hand-side of \eqref{eq:bincoeff}, entering the expression of $\Omega_N(M,2S)$, accounts for two specific configurations in which all spins are either up ($+1$) or down ($-1$). In such cases, the product of Kronecker delta's is equal to $1$ with the result that $\Omega_N(M=\pm N,2S=0)=1$.

Therefore, overall, the exact expression of the joint probability $P_N(M,2S)$ is:
\begin{eqnarray}
    P_N(M,2S) &=& \left[ \binom{\frac{N+M}{2}-1}{S-1} \binom{\frac{N-M}{2}}{S}
        +\binom{\frac{N-M}{2}-1}{S-1} \binom{\frac{N+M}{2}}{S}\right.\nonumber \\
    &+&\delta_{M,\pm N}\delta_{S,0}
    \Big] \frac{e^{\frac{J}{T}(N-4S) + \frac{K}{2NT}M^2 + \frac{h}{T} M}}{Z_N}.
\end{eqnarray}
The marginal probability distributions for the magnetization and the number of defects can be obtained by summing the joint distribution over the appropriate variable, i.e., $P_N(M) = \sum_{S=0}^{\lfloor N/2 \rfloor}P_N(M,2S)$ and $P_N(2S) = \sum_{M=-N}^{N} P_N(M,2S)$, where $\lfloor x \rfloor$ denotes the largest integer smaller or equal to $x$.

The expression of the probability distribution $P_N(M)$ can be rewritten in terms of hypergeometric functions. In particular, as detailed in Appendix \ref{app:JPDF}, $P_N(M)$ is equal to
\begin{eqnarray}\label{eq:Pnmhyp}
   P_N(M) &=& N \frac{e^{ \frac{J}{T} (N-4) + \frac{K}{2NT}M^2 + \frac{h}{T} M}}{Z_N}\nonumber\\
   &\times& \setlength\arraycolsep{1pt}
{}_2F_1\Big(1-\frac{N+M}{2},1-\frac{N-M}{2},2;e^{-4J/T}\Big),\quad
\end{eqnarray}
where ${}_2F_1(a,b,c;\zeta)$ denotes the hypergeometric function with parameters $a, b, c, \zeta$~\cite{abramowitz1965handbook}.

\section{Large-deviation behavior in the large-$N$ limit}\label{sec4}

In the large $N$ limit, the use of the Stirling formula $n! \sim \sqrt{2\pi n} {(n/e)}^n$ allows us to extract the dominant behavior of the joint distribution $P_N(M,2S)$ of magnetization and number of defects, which is provided by the following expression in large-deviation form~\cite{touchette2009large}: 
\begin{equation}
P_N(M,2S) \asymp e^{{-N U(m,s)}},
\end{equation}
where $\asymp$ means that $U(m,s) = -\lim_{N\rightarrow + \infty} \log\big(P_N(M,2S)\big)/N$. Here, $U(m,s)$ is a non-negative quasi-potential given by
\begin{eqnarray}\label{eq:potentialU}
    &U(m,s) = s \log (s)-\frac{1+m}{2} \log(1+m) -\frac{1-m}{2} \log(1-m)&\nonumber \\
    &+\frac{1+m-s}{2} \log(1+m-s) +\frac{1-m-s}{2} \log (1-m-s) &\nonumber \\
    &- \frac{1}{T}\left(  \frac{K}{2} m^2 + h m + J(1-2s) + f \right),
\end{eqnarray}
where $f$ is the shortcut notation for the system's free-energy. The defect density $s$ satisfies the bounds $0\leq s\leq 1-\vert m\vert$. This is because the magnetization $M$ takes values from $-N$ to $N$, in steps of $2$, while the defects number ranges from $2$ to $N-\vert M\vert$, also in steps of $2$ (except when $M=\pm N$, for which there is no defect). 

\begin{figure}[t]
    \centering
    \begin{subfigure}[t]{\linewidth}
        \centering
        \includegraphics[width=0.99\linewidth]{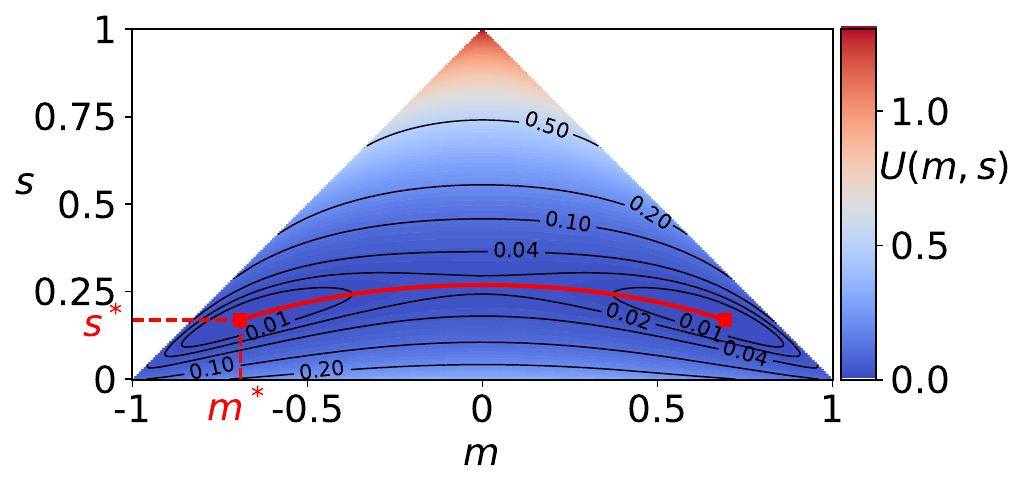}
        \label{fig:top_panel}
    \end{subfigure}
    \vspace{-0.5cm}
    \begin{subfigure}[t]{\linewidth}
        \centering
        \includegraphics[width=0.99\linewidth]{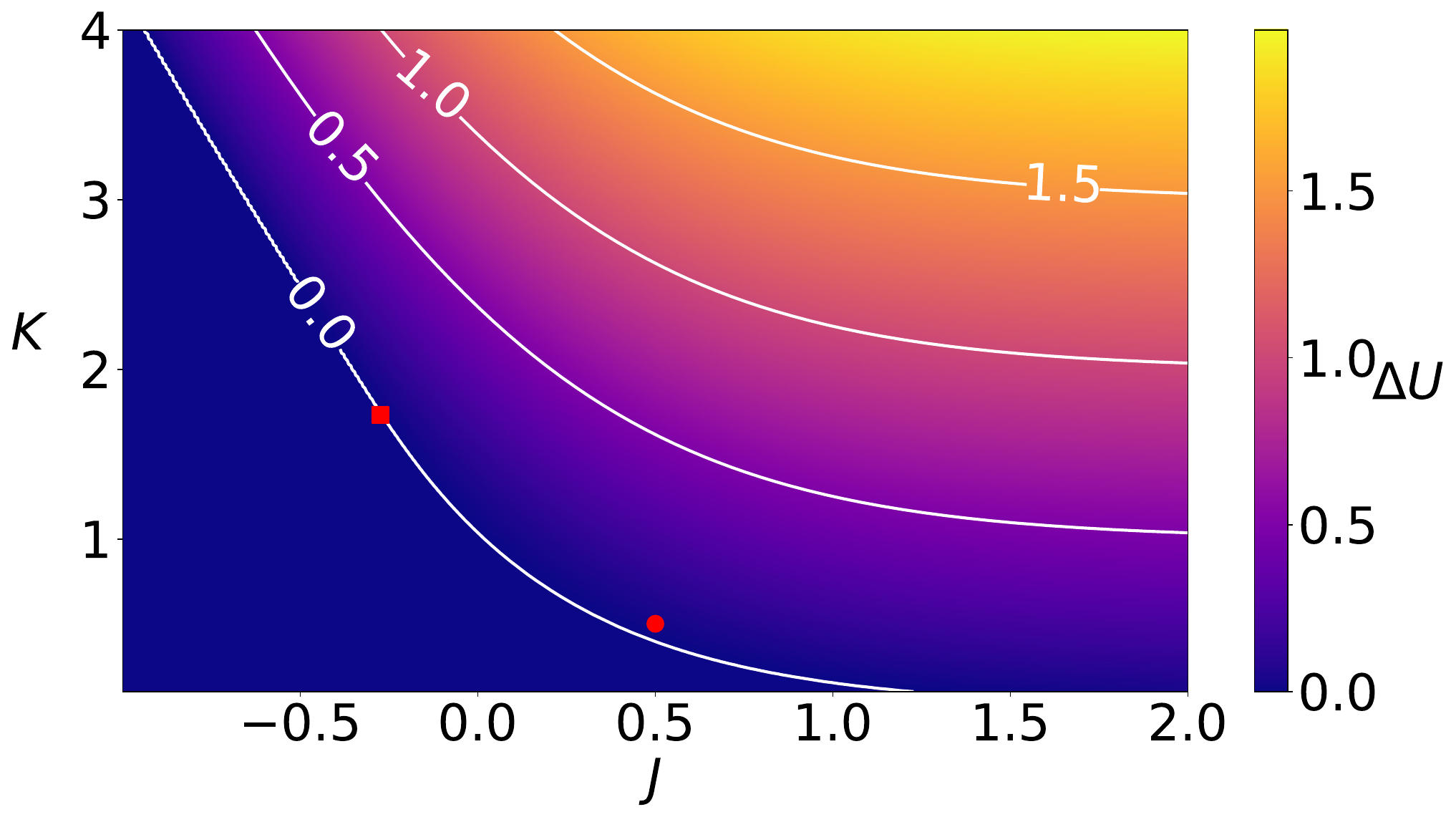}
        \label{fig:bottom_panel}
    \end{subfigure}
    \caption{
    (Top) Heat map of the Quasi-potential $U(m,s)$ with parameters $T=1$, $h=0$, $J=K=1/2$ . The two minima $(m^*,s^*)$, such that $U(m^*,s^*)=0$, are highlighted by the red squares. A few contour lines are shown in black. The red curve represents the solution of \eqref{eq:ssp}. (Bottom) Heat map of the Quasi-potential barrier $\Delta U$ for $h=0$ and $T=1$, separating the two local minima of $U(m,s)$, plotted as a function of $J$ and $K$, together with a few contour lines. The contour line $\Delta U=0$ marks the second-order phase transition for $K < \sqrt{3}$ and the first-order one for $K >\sqrt{3}$. The red circle at $J=K=0.5$ corresponds to the quasi-potential barrier $\Delta U \approx 0.018$ of the top panel, while the red square, where $\Delta U=0$, is the tricritical point $(J\!=\!-\frac{\log(3)}{4},K\!=\!\sqrt{3})$. 
    }
    \label{fig:2D_potential}
\end{figure}

In Fig.~\ref{fig:2D_potential} we plot the potential $U(m,s)$ for $h=0$; in absence of the external field the potential is symmetric in $m$. As $N\rightarrow + \infty$, the equilibrium values of the magnetization per site $m^*$ and fraction of defects $s^{*}$ are determined by the condition $U(m^*,s^*)=0$. The magnetization $m^*$ satisfies the following implicit equation~\cite{Dantchev2024}:
\begin{equation}
    m^{*}=\frac{\sinh\Big( (h+Km^*)/T \Big)}{\sqrt{\sinh^2\Big( (h+Km^*)/T \Big) + e^{-4 J/T}} }.
    \label{eq:msp}
\end{equation}
Instead, the fraction of defects $s^*$ depends on $m^*$:
\begin{equation}
    s^* = \frac{ 1 - \sqrt{{m^*}^2 + e^{4 J/T} (1 - {m^*}^2)}}{1-e^{4 J/T}}.
    \label{eq:ssp}
\end{equation}
In the strongly antiferromagnetic limit $J\rightarrow -\infty$, the spin system favors the perfectly alternating spin configuration, corresponding to $s^* = 1$. In contrast, when $J/T \rightarrow 0$, the magnetization per site is $m^*=\tanh( (h+Km^*)/T )$ and $s^*\approx\frac{1-{m^*}^2}{2}$.

The probability distribution of the magnetization, $P_N(M)$, can be obtained by taking the marginal of the distribution $P_N(M,2S)$, or using an asymptotic expression of the hypergeometric functions (see Appendix \ref{app:LD} for the derivation). One thus obtains: 
\begin{equation}
P_N(M) \asymp e^{-N \psi(m)},
\end{equation}
where 
\begin{eqnarray}\label{eq:LDVmagn}
    \psi(m) = &-& \frac{1}{T}\left( J - h m - \frac{K}{2}m^2 - f\right) - \frac{1+m}{2} \log\left( 1 - \frac{1}{\xi(m)} \right)\nonumber\\ 
    &-& \frac{1-m}{2} \log \left( 1 - e^{-4 J/T}\xi(m)\right),    
\end{eqnarray}
with $\xi(m) = -\frac{m}{1-m} - \frac{1}{1-m} \sqrt{m^2 + e^{4 J/T} (1-m^2)}$. The function $\psi(m)$ is non-negative and vanishes at $m=m^*$.

In the disordered phase and for $h=0$, the large-deviation function $\psi(m)$ admits the following expansion around $m=0$:
\begin{eqnarray}
    \psi(m) &=& \frac{e^{-2J/T} -\frac{K}{T}}{2} m^2 + \frac{3e^{4J/T}-1}{24 e^{6J/T}} m^4 \nonumber \\
    &+& \frac{15 - 10e^{-4J/T} + 3e^{-8 J/T}}{240 e^{2J/T}}m^6 +  O(m^8).
\end{eqnarray}
Away from the second-order transition line $K/T = e^{-2 J/T}$, $\psi(m)$ is quadratic around $m=0$. In contrast, on the line $K/T = e^{-2J/T}$, $\psi(m)$ has a quartic behavior around $m=m^*=0$, except at the tricritical point where the coefficient in $m^4$ also vanishes. From Eq.~(\ref{eq:LDVmagn}) we also deduce that $\vert m \vert \sim N^{-1/2}$ in the disordered phase and away from the second-order transition line, while $\vert m \vert \sim N^{-1/4}$ at the second-order line $K/T = e^{-2 J/T}$, except at the tricritical point where $\vert m \vert \sim N^{-1/6}$. We recover the results of \cite{gaspard2012fluctuation} in the limiting cases of the one-dimensional Ising ($K=0$) and Curie-Weiss ($J=0$) models.

\begin{figure}
    \centering
    \includegraphics[width=0.99\linewidth]{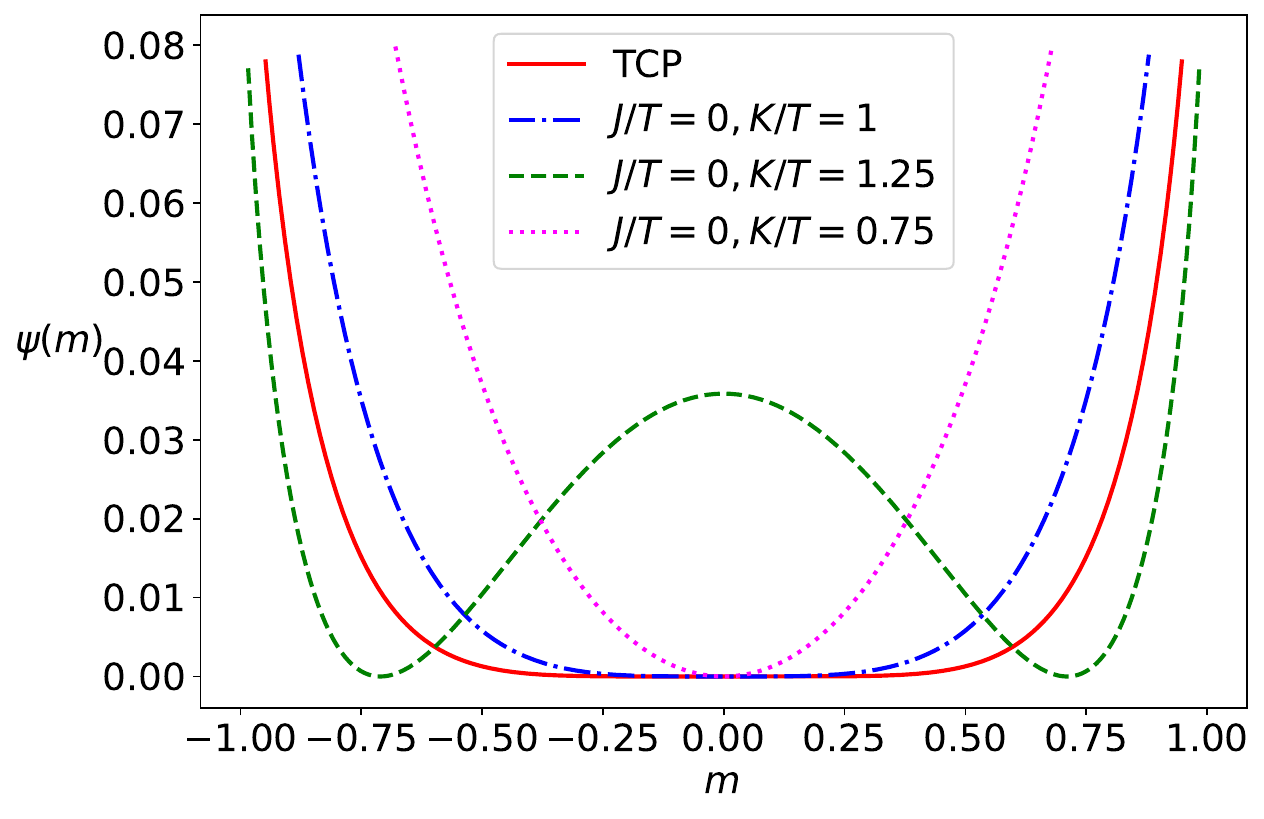}
    \caption{
    Plot of $\psi(m)$ \eqref{eq:LDVmagn} for several sets of parameters for $h=0$. Around $m=0$, $\psi(m) = O(m^2)$ in the disordered phase (dotted magenta), $\psi(m) = O(m^4)$ along the critical line (dash-dotted blue line), while $\psi(m) = O(m^6)$ at the tricritical point [TCP in the legend of the figure] (solid red line). We also plot $\psi(m)$ for $J=0$, $K=1.25$, $h=0$, $T=1$, which belongs to the ferromagnetic phase (green dashed line). 
    }
    \label{figSM:Large_deviation_magnetization}
\end{figure}

In Fig.~\ref{figSM:Large_deviation_magnetization}, we plot $\psi(m)$ for different values of the parameters across the second order phase transition. The magenta dotted curve, obtained for $J/T = 0$ and  $K/T=0.75$, corresponds to the disordered phase: it has a unique minimum for $m=m^*=0$, the equilibrium value of $m$, around which the function has a quadratic behavior. At the second-order transition line, $\psi(m)\sim m^4$ for small $m$ (dash-dotted blue curve). At the tricritical point, $\psi(m)\sim m^6$ around $m=0$ (full red line). In all these cases, the function $\psi(m)$ always vanishes at its minimum, meaning that, by increasing $N$, the probability distribution of the magnetization concentrates on its equilibrium value. The situation changes in the ferromagnetic phase (dashed green line), where the large deviation function vanishes at two symmetric minima where $m^*\neq 0$.

\section{Pair-correlation function}
\label{sec5}

Let us now turn to the computation of the {\it pair-correlation function} $\langle s_i s_j \rangle - \langle s_i \rangle \langle s_j \rangle$, with $\langle f\rangle := Z_N^{-1}\sum_{\boldsymbol{\sigma}} f(\boldsymbol{\sigma}) \, e^{- \frac{H(\boldsymbol{\sigma})}{T} }$, following the methods exposed in Baxter's book~\cite{baxter1985exactly}. We thus obtain:
\begin{eqnarray}\label{eq:pair_corr_func}
    &&\langle s_i s_j \rangle = \nonumber\\ 
    &=& \displaystyle{\sqrt{ \frac{NK}{2\pi T Z_N^2} }    \sum_{\boldsymbol{\sigma}} 
	\int_{-\infty}^{+\infty} dx \, s_i s_j \,  e^{- \frac{N K}{2 T} x^2 +  \frac{J}{T}\sum_{i=1}^N s_i s_{i+1} + \frac{1}{T} (Kx+h) \sum_{i=1}^{N} s_i} }\nonumber\\
    &=& \displaystyle{ \sqrt{ \frac{NK}{2\pi T Z_N^2} }
		\int_{-\infty}^{+\infty}dx\,e^{- \frac{N K}{2 T} x^2}\, \text{Tr}\left( SV^{j-i}SV^{N-j+i} \right),}
\end{eqnarray}
where we used the invariance under translation assuming $0\leq j-i\leq \lfloor N/2 \rfloor$. This assumption comes from the consideration that, given two lattice sites $i$ and $j>i$, the distance $r$ between them is the minimum between $(j-i)$ and $(i-j) + N$, with $0 \leq r \leq \lfloor N/2 \rfloor$. Moreover, in \eqref{eq:pair_corr_func},
$S=\begin{pmatrix}
    1 & 0\\
    0 &-1
\end{pmatrix}$, and $V$ is the transfer matrix of \eqref{eq:Trmat}. The matrix $V$, being symmetric, is diagonalizable by an orthogonal matrix $P$, such that
\begin{equation}
    P^{-1}VP = 
    \begin{pmatrix}
    \lambda_+ & 0\\
    0 &\lambda_{-}
\end{pmatrix} := \widetilde{V},
\end{equation}
where the eigenvalues $\lambda_{\pm}$ are those given in~\eqref{eq:lambda}. The matrix $P$ can be written as 
\begin{equation}
P=\begin{pmatrix}
    \cos(\phi) & -\sin(\phi)\\
    \sin(\phi) & \cos(\phi)
\end{pmatrix},
\end{equation}
with $\phi\equiv \phi(x)$ solution of the equation $\cot\left( 2\phi(x) \right) = e^{2J/T} \sinh \left( (h+Kx)/T \right)$. It follows that 
\begin{eqnarray}\label{eq:pair_corr_func_2}
    \langle s_i s_j\rangle &=& \sqrt{\frac{NK}{2\pi T Z_N^2}}\times\nonumber\\
    &\times& \int_{-\infty}^{+\infty}dx\,e^{- \frac{NK}{2T} x^2}\, \text{Tr}\left( P^{-1}SP\widetilde{V}^{j-i}P^{-1}SP\widetilde{V}^{N-j+i} \right).\qquad
\end{eqnarray}

We first evaluate the trace of the product of matrices on the right-hand-side of \eqref{eq:pair_corr_func_2}, which is given by
\begin{equation}
\cos^{2}(2\phi) \, \left( \lambda_+^{N}+\lambda_{-}^{N} \right) + \sin^{2}(2\phi) \left( \lambda_{+}^N \left( \frac{\lambda_{-}}{\lambda_{+}} \right)^{j-i}
+\lambda_{-}^N \left( \frac{\lambda_{+}}{\lambda_{-}} \right)^{j-i}
\right).\nonumber 
\end{equation}
As we are interested in the large $N$ limit, we perform a saddle-point analysis. Let us first note that, for large $N$,
\begin{eqnarray}
&&\lambda_+^{N}\left[
\cos^{2}(2\phi) \, \left( 1+{\left( \frac{\lambda_{-}}{\lambda_{+}}\right)}^{N}\right) + \sin^{2}(2\phi) 
\left( {\left(\frac{\lambda_{-}}{\lambda_{+}}\right)}^{r}
+{\left(\frac{\lambda_{-}}{\lambda_{+}}\right)}^{N} {\left(\frac{\lambda_{+}}{\lambda_{-}}\right)}^{r}
\right)\right] \nonumber \\
&&\sim e^{N \log\left( \lambda_{+} \right)} \left[
\cos^{2}(2\phi) + \sin^{2}(2\phi) \left(\frac{\lambda_{-}}{\lambda_{+}}\right)^{r} \right],
\end{eqnarray}
where $\lambda_{-} \leq \lambda_{+}$. Hence, for large $N$, we find:
\begin{eqnarray}\label{eq:sisj}
&&\langle s_i s_j \rangle \sim\nonumber\\
&\sim& \frac{
\displaystyle{
\int_{-\infty}^{+\infty} dx \, e^{NF(x)} \left( \cos^{2}\big( 2\phi(x) \big) + \sin^{2}\big( 2\phi(x) \big) \left( \frac{\lambda_{-}(x)}{\lambda_{+}(x)}\right)^{r}\,\right) }
}
{
\displaystyle{
\int_{-\infty}^{+\infty} dx \, e^{NF(x)}
}},\nonumber\\
\end{eqnarray}
where $F(x) := - \frac{K}{2 T} x^2 +\log\left( \lambda_{+}(x) \right)$. A similar computation leads us to
\begin{equation}
\langle s_i \rangle \sim 
\frac{
\int_{-\infty}^{+\infty}dx \, \cos(2\phi(x)) \, e^{N F(x)}
}
{
\int_{-\infty}^{+\infty} dx \, e^{N F(x)}
}.
\end{equation}
Overall, the pair-correlation function can be written as
\begin{eqnarray}\label{eq:gr1}
    g(r) := 
    \langle s_{i}s_{i+r}\rangle - \langle s_i\rangle^{2}.
\end{eqnarray}

We now focus on the asymptotic behavior of $g(r)$ in $1/N$ along the critical line, in the $h=0$ plane. In this case, $\langle s_i \rangle = 0$ and we find (see also Appendix~\ref{app:gr}):
\begin{equation}\label{eq:asymg}
g(r) \sim \left( \frac{1-K}{1+K} \right)^r + \frac{\Gamma(3/4)}{\Gamma(1/4)} \sqrt{ \frac{24}{K(3-K^2)}} \frac{1}{\sqrt{N}}  
\end{equation}
for $0<K<\sqrt{3}$, and
\begin{equation}\label{eq:asymg_2}
    g(r) \sim \left( \sqrt{3}-2 \right)^r + \frac{\Gamma(1/2)}{\Gamma(1/6)} \left( \frac{\sqrt{3}}{20} \right)^{-1/3} N^{-1/3}
\end{equation}
for $K=\sqrt{3}$. The pair-correlation function $g(r)$ always decays exponentially at large distances $r$ as shown by the behavior of first term in Eqs.~(\ref{eq:asymg})-(\ref{eq:asymg_2}). Therefore, the correlation length $\xi$, defined as $g(r) \underset{r \gg 1}{\sim} \exp (-r/\xi)$, is always finite and diverges exponentially only in the limit $(J \to \infty, K\to 0)$, the zero-temperature limit of the one-dimensional Ising model. However, the finite-size corrections have different exponents, whether the system is at the critical line (exponent $-1/2$) or at the tricritical point (exponent $-1/3$). Let us remark that the first term on the right-hand-side of the asymptotic expression for the Curie-Weiss model ($K=1$) vanishes exactly, which is compatible with the well-known fact that the correlation function is exactly $0$ in the thermodynamic limit.

\begin{figure}[t!]
    \centering
    \includegraphics[width=0.99\linewidth]{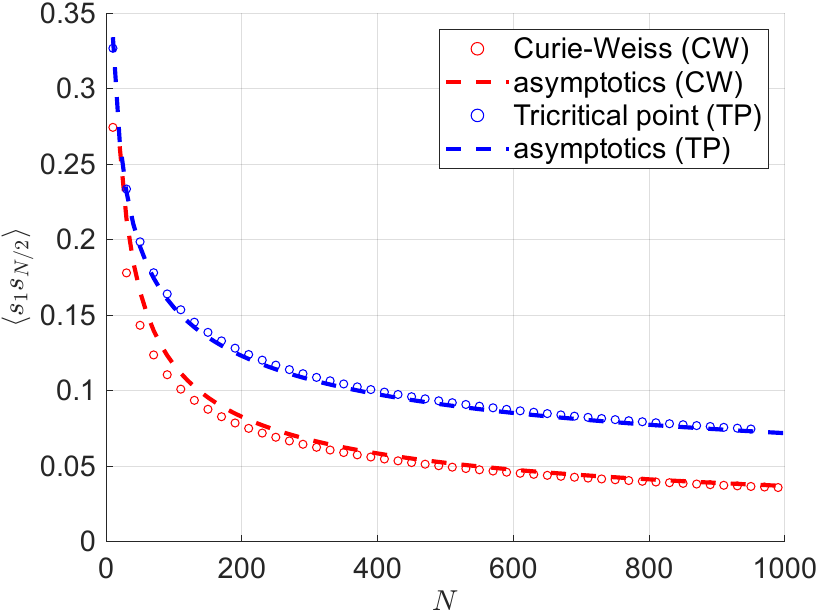}
    \caption{Pair-correlation function $g(r)$ for the Curie-Weiss model (dashed red line and dots) and for the tricritical point of the Nagle-Kardar model (dashed blue line and dots). The numerical values (dots) agree with the asymptotics (dashed lines) at large $N$.}
    \label{fig:gr}
\end{figure}

\begin{figure}[t!]
    \centering
    \includegraphics[width=0.99\linewidth]{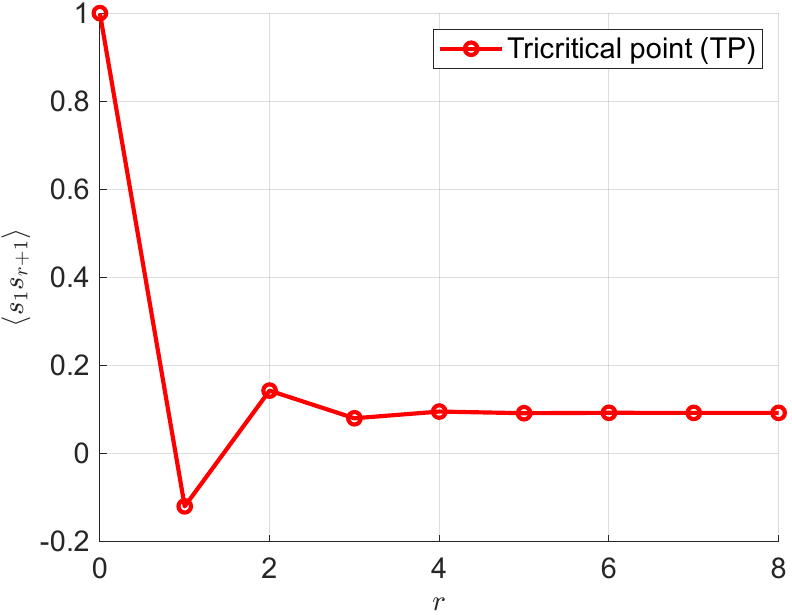}
    \caption{
    Pair-correlation function $\langle s_1 s_{r}\rangle$ at the tricritical point of the Nagle-Kardar model for $N=500$.
    }
    \label{fig:grTP}
\end{figure}

In Fig.~\ref{fig:gr}, we plot the pair-correlation function $g(r)$, \eqref{eq:gr1}, and asymptotic expressions, Eqs.~(\ref{eq:asymg})-(\ref{eq:asymg_2}), as a function of $N$. The value of $r$ is chosen as $r=N/2$.
In Fig.~\ref{fig:grTP}, we show $g(r)$, \eqref{eq:gr1}, at fixed $N$ as a function of $r$ for the tricritical point of the Nagle-Kardar model. For small $r$, the function displays oscillations due to the antiferromagnetic local coupling, $J<0$. For large $r$, the function decays exponentially to an $N$-dependent finite value which then vanishes in the infinite $N$ limit. The small-$r$ oscillations are rapidly damped because the behavior of the function $g(r)$ is dominated by the mean-field term in the Hamiltonian.

\section{Static and dynamical critical exponents}
\label{sec6}

In the companion Letter~\cite{deKemmeterLetter2025}, we have used the values of the static exponents $\beta$ and $\nu$ in the relations involving the finite-time and finite-size exponents
$\Delta_m$ and $\lambda_m$ at criticality:
\begin{equation}\label{eq:scaling_relations}
    \left\langle \left\vert m(t) \right\vert \right\rangle \sim t^{-\lambda_m} \quad \text{and} \quad \left\langle \left\vert m \right\vert \right\rangle \underset{t \to \infty}{\sim}  N^{-\Delta_m}.
\end{equation}
In this Section we derive the values of $\beta$ and $\nu$ at the critical line and at the tricritical point. For the exponent $\beta$ we consider the thermodynamic ($N \to \infty$) and static ($t \to \infty$) limits and we Taylor expand the magnetization $\langle \vert m\vert\rangle$ near criticality (Subsection \ref{beta}). As far as the exponent $\nu$ is concerned, we will resort to a finite-size and/or finite-time analysis.

\subsection{Exponent $\beta$}
\label{beta}

Let us consider~\eqref{eq:msp}, assuming $h=0$ and, without loss of generality, $T=1$:
\begin{equation}\label{eq:mh0}
    m^{*}=\frac{\sinh (Km^*) }{\sqrt{\sinh^2 (Km^*)  + K^2} }~.
\end{equation}
Then, we expand the right-hand-side of \eqref{eq:mh0} up to fifth-order around $m^*=0$, leading to
\begin{eqnarray}
    m^{*} = &&\frac{K m^*}{e^{-2J}} + \frac{(1-3e^{4J})K^3{m^*}^3}{6 \, e^{-2J}}\nonumber \\
    &&+ \frac{(-30 e^{4J} + 45e^{8J}+1)K^5 {m^*}^5}{120 \, e^{-2J}} + O\left( {m^*}^7 \right).
\end{eqnarray}
We denote by $\Delta$ the infinitesimal relative distance of the parameter $K$ from the critical line, i.e., $\Delta := (K-K_c)/K_c$ (thus, $K = K_c(1+\Delta)$) at fixed $J$. Setting the value of $K_c$ on the critical line ($K_c = e^{-2J}$), we get:
\begin{eqnarray}\label{eq:taylor}
    m^{*} &=& (1+\Delta) m^* - \frac{(3-K_c^2){(1+\Delta)}^3{m^*}^3}{6 }\nonumber\\
    &-& \frac{(-K_c^4 + 30 K_c^2 - 45){(1+\Delta)}^5 {m^*}^5}{120} + O\left( {m^*}^7 \right).
\end{eqnarray}
Besides the solution for vanishing magnetization $m^*=0$, \eqref{eq:taylor} has a non-zero solution of $m^*$ that scales like $m^* \sim \Delta^{\beta}$ with $\beta=1/2$ for $K_c^2 < 3$. At $K_c^2=3$, the cubic term in $m^*$ of the right-hand-side of \eqref{eq:taylor} vanishes and the fifth-order term is non-zero. Therefore, $m^*$ scales like $\Delta^\beta$ with $\beta=1/4$ at the tricritical point.

\subsection{Exponent $\nu$}
\label{nu}

The exponent $\nu$ characterizes the divergence of the correlation length near the critical line. In Section \ref{sec5}, we have shown that in the Nagle-Kardar model the correlation function cannot diverge with an inverse power-law near the critical line. This is also true both for the Curie-Weiss model ($J=0)$, whose pair correlation function is identically zero in the thermodynamic limit, and for the one-dimensional Ising model ($K=0$).

In order to define $\nu$ for the Nagle-Kardar model, one has to resort to finite-size scaling theory~\cite{FisherPRL1972,Barber1983,Privman1990Editor,wang1995study,ma2018modern}, for which the correlation length is replaced by the linear size of the system, $\xi \sim L$ ($L=N$ in $d=1$). Denoting as $\Delta$ the relative distance of one of the parameters of the Hamiltonian from the critical line, e.g., $\Delta = (K-K_c)/K_c$ at fixed $J$, the exponent $\nu$ is defined by the relation $\xi \sim L \sim |\Delta|^{\nu}$. Since $\langle\vert m\vert\rangle$ vanishes as $\langle\vert m\vert\rangle \sim |\Delta|^\beta$ near the critical line, one gets $\langle\vert m\vert\rangle \sim L^{-\beta/\nu}$, or, in dimension $d$ ($L^d=N$), that $\langle\vert m\vert\rangle \sim N^{-\beta/(\nu d)}$. This latter relation allows to derive the exponent $\Delta_m = \beta/(\nu d)$. In Section~\ref{sec4} we have shown that $\Delta_m=1/4$ at the critical line and $\Delta_m=1/6$ at the tricritical point. Moreover, we have just proved in Section \ref{beta} that $\beta=1/2$ at the critical line and $\beta=1/4$ at the tricritical point.

Both from probabilistic~\cite{Ellis1978,Ellis1980} and field-theory~\cite{Kenna2004,Honchar2024} arguments, one can generalize the relation that defines the exponent $\Delta_m$ in finite dimension $d$ to mean-field models, by replacing $d$ with the upper critical dimension $d_u$:
\begin{equation}
\label{eq:new_scaling_law}
\Delta_m = \frac{ \beta }{ \nu \, d_{u} }.
\end{equation}
In the companion Letter~\cite{deKemmeterLetter2025}, using the Ginzburg criterion, we have shown that $d_u=4$ on the critical line and $d_u=3$ at the tricritical point. From~\eqref{eq:new_scaling_law} and the previous determination of $\beta$, $d_u$ and $\Delta_m$, it directly follows that $\nu=1/2$. Let us remark that the value of $\nu$ does not depend on whether the system is on the critical line or at the tricritical point. This fact has a physical meaning in the mean-field Landau-Ginzburg theory of Ising-type models. Being critical or tricritical modifies only the Landau potential $\psi(m)$, but leaves unchanged the kinetic energy part of the action. Indeed, the calculation of the Ornstein-Zernike $g(r)$ in the Landau-Ginzburg theory involves only the kinetic energy~\cite{Goldenfeld1992Lectures}.

This derivation of the exponent $\nu$ is compatible also with a similar scaling theory based on the exponent $\lambda_m$, whose value is $1/2$ on the critical line and $1/4$ at the tricritical point. Using the Hohenberg-Halperin dynamical scaling theory~\cite{hohenberg1977theory}, the correlation length $\xi$ can be related to the correlation time $t$ by the scaling relation $\xi \sim t^{1/z}$. Since $\langle\vert m\vert\rangle$ scales as $\xi^{-\beta/\nu}$, at finite-size $\xi \sim L$, one gets $\langle\vert m\vert\rangle \sim t^{-\beta/(\nu z)}$, which implies
\begin{equation}
\label{eq:new_scaling_law_time}
\lambda_m = \frac{ \beta }{ \nu \, z }.
\end{equation}
The value $1/2$ of $\nu$, obtained above using the finite-size scaling theory, is also compatible with \eqref{eq:new_scaling_law_time} both at the critical line and at the tricritical point once one 
identifies the dynamical exponent $z=2$~\cite{deKemmeterLetter2025}.

\section{Glauber dynamics and master equation}\label{sec7}

In this section we provide the formal derivation of the master equation that governs the time evolution of the joint probability $P_N(M,2S;t)$ under Glauber dynamics. We recall that the number $2S$ of defects separating two adjacent spins with opposite signs is always even.

The Glauber dynamics proceeds as follows: starting from an initial configuration, a single spin is randomly selected at each discrete time step, and a flip is attempted. Denoting with $\Delta E$ the energy change associated with this flip, the move is accepted with probability ${(1+e^{\Delta E/T})}^{-1}$.
Upon acceptance of a single flip, the total number of up-spins increases or decreases by $2$; in the same way, the number of defects remains unchanged or changes by $\pm2$. Let $\mathcal{C}=(s_1,\cdots,s_N)$ be one of the $2^N$ distinct spin configurations. Then, the probability of observing a configuration $\mathcal{C}$ evolves in time according to the following master equation:
\begin{eqnarray}\label{SMeq:ME}
    \frac{\partial P(\mathcal{C},t)}{\partial t} = &-&\sum_i^N \mathcal{T}_{M\rightarrow M-2s_i}^{2S\rightarrow 2S}(s_i\rightarrow -s_i\vert \mathcal{C}) P(\mathcal{C},t)
    \nonumber\\
    &-& \sum_i^N \mathcal{T}_{M\rightarrow M-2s_i}^{2S\rightarrow 2S+2}(s_i\rightarrow -s_i\vert \mathcal{C}) P(\mathcal{C},t)
    \nonumber\\
    &-& \sum_i^N \mathcal{T}_{M\rightarrow M-2s_i}^{2S\rightarrow 2S-2}(s_i\rightarrow -s_i\vert \mathcal{C}) P(\mathcal{C},t)
    \nonumber\\
    &+&\sum_i^N \mathcal{T}_{M+2s_i\rightarrow M}^{2S\rightarrow 2S}(s_i\rightarrow -s_i\vert \mathcal{C'}) P(\mathcal{C'},t) 
    \nonumber\\
    &+& \sum_i^N \mathcal{T}_{M+2s_i\rightarrow M}^{2S+2\rightarrow 2S}(s_i\rightarrow -s_i\vert \mathcal{C'}) P(\mathcal{C'},t)\nonumber\\
    &+& \sum_i^N \mathcal{T}_{M+2s_i\rightarrow M}^{2S-2\rightarrow 2S}(s_i\rightarrow -s_i\vert \mathcal{C'}) P(\mathcal{C'},t).
\end{eqnarray}
In \eqref{SMeq:ME}, the transition rates are decomposed as follows:
\begin{equation}\label{eq:transition_rate}
    \mathcal{T}_{M\rightarrow M'}^{2S\rightarrow 2S'}(s_i \rightarrow-s_i\vert\mathcal{C}) = I_{M\rightarrow M'}^{2S\rightarrow 2S'}(s_i \rightarrow-s_i\vert\mathcal{C}) \, A_{M\rightarrow M'}^{2S \rightarrow 2S'}, 
\end{equation}
where $A_{M\rightarrow M'}^{2S \rightarrow 2S'}$ denotes the acceptance probability of a spin flip, which changes the magnetization from $M$ to $M'$ and the number of defects from $2S$ to $2S'$. This probability depends solely on the associated energy change, and not on the specific configuration. Moreover, in \eqref{SMeq:ME}, $I_{M\rightarrow M'}^{2S\rightarrow 2S'}(s_i \rightarrow-s_i\vert\mathcal{C})$ is an indicator function that takes the value $1$ if the flipping site $i$ in the configuration $\mathcal{C}$ changes the magnetization from $M$ to $M'$ and the number of defects from $2S$ to $2S'$.

In order to reduce the complexity of the analysis, we now restrict our attention to the time-evolution of the probability $P_N(M,2S;t)$ at time $t$. The latter is given by
\begin{equation}\label{eq:ME2}
    P_N(M,2S;t) = \sum_{\mathcal{C}} \delta\left( \sum_i s_i, M\right) \delta\left( \sum_i \frac{1-s_i s_{i+1}}{2}, 2S\right) P(\mathcal{C},t),    
\end{equation}
where the sum runs over the $2^N$ distinct spin configurations $\mathcal{C}$. Using \eqref{SMeq:ME}, the time evolution of $P_N(M,2S;t)$ is:
\begin{eqnarray}\label{SMeq: dPNdt}
    \frac{\partial P_N(M,2S;t)}{\partial t} =&& -\sum_{\mathcal{C}} \delta\left(\sum_i s_i,M\right) \delta\left( \sum_i \frac{1-s_i s_{i+1}}{2},2S\right)\nonumber\\
    \times&&\left[ \sum_{i=1}^{N} \mathcal{T}_{M\rightarrow M-2}^{2S \rightarrow 2S}(s_i=+1\rightarrow s_i=-1\vert \mathcal{C}) P(\mathcal{C},t)\right.\nonumber\\
    &&+\sum_{i=1}^N \mathcal{T}_{M\rightarrow M+2}^{2S \rightarrow 2S}(s_i=-1\rightarrow s_i=+1\vert \mathcal{C}) P(\mathcal{C},t)\nonumber\\
    &&+\text{the other terms entering~\eqref{SMeq:ME}}\Big].    
\end{eqnarray}

Let us consider the first term in \eqref{SMeq: dPNdt}. The transition rate $\mathcal{T}_{M\rightarrow M-2}^{2S \rightarrow 2S}(s_i=+1\rightarrow s_i=-1\vert \mathcal{C})$ can be decomposed as a product of two terms [\eqref{eq:transition_rate}]; in order to obtain a closed expression for the time evolution of $P_N(M,2S;t)$, one has to get rid off the explicit (a priori) dependence of the indicator function $I_{M\rightarrow M-2}^{2S \rightarrow 2S}(s_i=+1\rightarrow s_i=-1\vert \mathcal{C})$ on the microscopic configurations $\mathcal{C}$. This indicator function is equal to $1$ if the spin $s_i$, in configuration $\mathcal{C}$ with magnetization $M$ and $2S$ defects, is $+1$ and adjacent to a defect. Thus, for the first term in \eqref{SMeq: dPNdt}, we have:
\begin{eqnarray}
&&\sum_{\mathcal{C}}
\delta\!\left( \sum_i s_i , M\right)
\delta\!\left(\sum_i \tfrac{1-s_i s_{i+1}}{2}, 2S\right)
P(\mathcal{C},t)\nonumber\\
&&\times
\sum_{i=1}^N
\mathcal{T}_{M\!\rightarrow\! M-2}^{\,2S\!\rightarrow\! 2S}
(s_i{=}{+}1\!\rightarrow\! s_i{=}{-}1 \mid \mathcal{C})\nonumber \\
&&=A_{M\!\rightarrow\! M-2}^{\,2S\!\rightarrow\! 2S}
\sum_{k\ge0} k\,\widetilde{P}_N(M,2S,k;t),
\end{eqnarray}
where $\widetilde{P}_N(M,2S,k;t)$ denotes the probability to observe a configuration with magnetization $M$, $2S$ defects and $k$ up-spins surrounded by a single defect at time $t$. Then, we consider the definition of conditional probability:
\begin{eqnarray}
    \widetilde{P}_N(M,2S,k;t) &=& P_N(k\vert M,2S;t) P_N(M,2S;t)\nonumber\\
    &\approx& \frac{\Omega_1^+(N,M,2S,k)}{\Omega_N(M,2S)}P_N(M,2S;t),
\end{eqnarray}
where we have used the assumption that the system explores all microstates with equal probability, once fixed the values of $N$, $M$ and $2S$, at any time $t$. Such assumption allows us to replace the time-dependent conditional probability $P_N(k\vert M,2S;t)$ with a time-independent combinatorial factor. Specifically, $\Omega_1^+(N,M,2S,k)$ is the number of all possible configurations with magnetization $M$, $2S$ defects and $k$ up-spins (superscript $+$) surrounded by a single defect (subscript $1$), while $\Omega_N(M,2S)$ is defined in \eqref{eq:bincoeff}.

\begin{figure}
\centering
\includegraphics[width=0.99\linewidth]{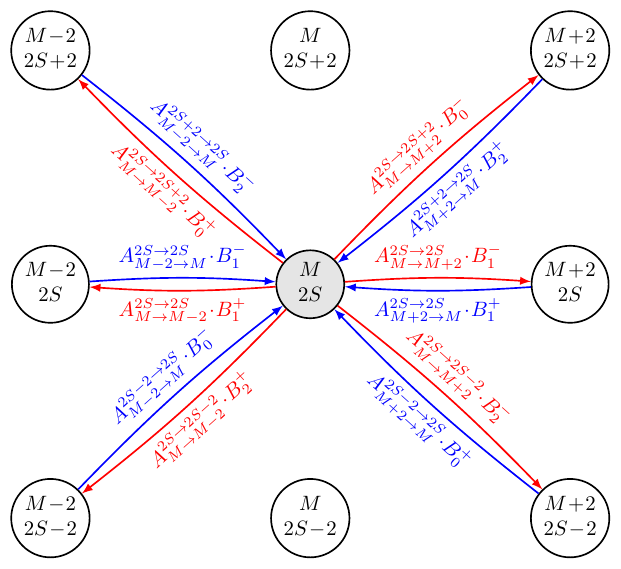}
    \caption{
    Allowed transition rates from and to the macrostate $(M,2S)$, following \eqref{SMeq:dPNdt2}.
    }
\label{fig:transrates}
\end{figure}

Proceeding in the same manner for the other terms, we can obtain a closed expression for $\frac{\partial}{\partial t}P_N(M,2S;t)$:
\begin{widetext}
\begin{eqnarray}\label{SMeq:dPNdt2}
    &&\frac{\partial P_N(M,2S;t)}{\partial t}=\nonumber \\ 
    && = - A_{M\rightarrow M-2}^{2S\rightarrow 2S}  
    B_1^{+}(N,M,2S)P_N(M,2S;t) 
    - A_{M\rightarrow M+2}^{2S\rightarrow 2S} 
    B_1^{-}(N,M,2S) P_N(M,2S;t) -A_{M\rightarrow M-2}^{2S\rightarrow 2S+2} 
    B_0^{+}(N,M,2S)P_N(M,2S;t)\nonumber\\
    && + A_{M\rightarrow M+2}^{2S\rightarrow 2S+2} 
    B_0^{-}(N,M,2S) P_N(M,2S;t) - A_{M\rightarrow M-2}^{2S\rightarrow 2S-2} 
    B_2^{+}(N,M,2S)P_N(M,2S;t) + A_{M\rightarrow M+2}^{2S\rightarrow 2S-2} 
    B_2^{-}(N,M,2S) P_N(M,2S;t)\nonumber\\
    && + A_{M+2\rightarrow M}^{2S\rightarrow 2S} 
    B_1^{+}(N,M+2,2S)P_N(M+2,2S;t) + A_{M-2\rightarrow M}^{2S\rightarrow 2S} 
    B_1^{-}(N,M-2,2S) P_N(M-2,2S;t)\nonumber\\ 
    && + A_{M+2\rightarrow M}^{2S+2\rightarrow 2S} 
    B_2^{+}(N,M+2,2S+2)P_N(M+2,2S+2;t) + A_{M-2\rightarrow M}^{2S+2\rightarrow 2S} 
    B_2^{-}(N,M-2,2S+2) P_N(M-2,2S+2;t)\nonumber\\
    && + A_{M+2\rightarrow M}^{2S-2\rightarrow 2S} 
    B_0^{+}(N,M\!+\!2,2S\!-\!2) P_N(M\!+\!2,2S\!-\!2;t) + A_{M-2\rightarrow M}^{2S-2\rightarrow 2S} 
    B_0^{-}(N,M\!-\!2,2S\!-\!2) P_N(M\!-\!2,2S\!-\!2;t)
\end{eqnarray}
\end{widetext}
with
\begin{eqnarray}
    B_i^{\pm}(N,M,2S) & := & \left( \sum_{\ell \geq 0} \ell \frac{\Omega_i^{\pm}(N,M,2S,\ell)}{\Omega_N(M,2S)} \right).
\end{eqnarray}
For clarity, all the transitions rates entering \eqref{SMeq:dPNdt2} are depicted in Fig.~\ref{fig:transrates}.

In the following subsections, we are going to determine the explicit expressions of both the acceptance probabilities $A_{M\rightarrow M'}^{2S \rightarrow 2S'}$, and the combinatorial factors $B_i^{\pm}(N,M,2S)$.

\subsection{Acceptance probabilities}

Let $\Delta E = E_{\text{new}} - E_{\text{old}}$ denote the energy change resulting from a transition between a configuration of energy $E_{\text{old}}$ to one with energy $E_{\text{new}}$. Thus, the acceptance probability under the Glauber dynamics is given by ${(1+e^{\Delta E/T})}^{-1}$. We distinguish three cases depending on how the number of defects is affected by the single flip.

First we consider the case in which flipping the spin $i$ leaves unchanged the number of defects, i.e., $M\rightarrow M-2s_i$ and $2S \rightarrow 2S$. In such a case, the energy of the new and old configurations are respectively:
\begin{equation}
    \begin{split}
        & E_{\text{new}} = -J(N-4S) - \frac{K}{2N}{(M-2s_i)}^2 - h(M-2s_i),\\
        & E_{\text{old}} = -J(N-4S) - \frac{K}{2N}M^2 -hM.
    \end{split}
\end{equation}
Hence, the resulting acceptance probability reads as
\begin{equation}\label{eq:Glauber1}
        A_{M\rightarrow M-2s_i}^{2S\rightarrow 2S} = \frac{
        e^{-s_i [K (M-s_i)/N + h]/T}
        }{
        2\cosh \{ [K(M-s_i)/N + h]/T \}
        }.    
\end{equation}
Second, for the case in which flipping the spin $i$ reduces the number of defects by $2$, i.e., $M\rightarrow M-2s_i$ and $2S \rightarrow 2S-2$, the acceptance probability is given by
\begin{equation}
        A_{M\rightarrow M-2s_i}^{2S\rightarrow 2S-2} = \frac{
        e^{ - s_i [K(M-s_i)/N + h -2J s_i]/T}
        }{
        2\cosh \{ [K(M-s_i)/N + h -2J s_i]/T \}
        }.
\end{equation}
Finally, we consider the case where flipping the spin $i$ increases the number of defects by $2$, i.e., $M\rightarrow M-2s_i$ and $2S \rightarrow 2S+2$. The acceptance probability for this local move is:
\begin{equation}
        A_{M\rightarrow M-2s_i}^{2S\rightarrow 2S+2} = \frac{
        e^{ - s_i [K(M-s_i)/N + h + 2J s_i]/T }
        }{
        2\cosh \{ [K(M-s_i)/N + h + 2J s_i]/T \}
        }.
\end{equation}

\subsection{Combinatorial enumerations}

We now turn our attention to the computation of the coefficients $\Omega_i^{\pm}(N,M,2S,k)$ that enumerate configurations of $N$ spins with magnetization $M$, $2S$ defects and $k$ spins $\pm 1$ surrounded by $i$ defects ($i=0,1,2$). We start with the enumeration of up-spins ($+1$) surrounded by a single defect.

\subsubsection{Enumeration of up-spins adjacent to a single defect}

The number of up-spins adjacent to a single defect must be even. To see this, image to remove all spins $+1$ that are surrounded by two defects. This leaves an even number of defects, each paired with exactly one remaining spin $+1$. Since now no pair of two defects shares the same spin $+1$ (such defects indeed have been removed), the total number of such spins must be even. 

The number $\Omega^+_1(N,M,2S,2k)$ of configurations of $N$ spins, magnetization $M$, $2S$ defects and $2k$ up-spins adjacent to a single defect is:
\begin{equation}
    \Omega^+_1(N,M,2S,2k) = \left\{\begin{array}{ll}
    \frac{2N}{N-M}\binom{\frac{N-M}{2}}{\frac{N+M}{2}} \quad\text{ if } S=\frac{N+M}{2} \text{ and } k=0,\\[2mm]
    \frac{2N}{N-M} \binom{\frac{N-M}{2}-k}{S-k} \binom{\frac{N+M}{2}-S-1}{k-1}\binom{\frac{N-M}{2}}{k}  \text{ otherwise}.
    \end{array}\right.
    \label{SMeq:conj1}
\end{equation}
\eqref{SMeq:conj1} is valid for any $N$ and $M$, provided they have the same parity (i.e., $M = -N, -N+2, \dots, N-2, N$); otherwise, some binomial coefficients would involve half-integers, yielding zero. Moreover, \eqref{SMeq:conj1} holds if $M<N$ and $k\geq 1$. When $M=N$, all spins are $+1$ and there is no defect. We thus have $\Omega^+_1(N,N,0,0) = 1$.

The number $\Omega^+_1(N,M,2S,2k)$ grows exponentially with $N$. To study its growth rate, let us introduce the intensive (rescaled by $N$) variables
\begin{equation}\label{eq:intensive_variables}
    m := \frac{M}{N}, \quad \kappa := \frac{2k}{N}, \quad s := \frac{2S}{N}.
\end{equation}
We thus have:
\begin{equation}
    \lim_{N\rightarrow + \infty} \frac{\log \left( \Omega^+_1(N,M,2S,2k) \right)}{N} = \frac{1}{2}f_1(m,s,\kappa),
\end{equation}
where
\begin{eqnarray}
        &f_1(m,s,\kappa) =
        -(s-\kappa)\log (s-\kappa) - (1-m-s) \log (1-m-s)\nonumber\\
        &+ (1-m)\log (1-m)+ (1-s+m)\log(1+m-s)\nonumber\\
        &-2\kappa \log(\kappa) - (1-s + m - \kappa)\log(1-s+m - \kappa),
\end{eqnarray}
which represents the growth rate of $\Omega^+_1$ in the large-$N$ limit: $\Omega^+_1 \asymp e^{Nf_1/2}$. In such a limit ($N$ large), main contribution to $\Omega_1^+(N,M,2S,2k)$ is achieved in correspondence (and in the neighborhood) of ${\kappa}_{1}^+$ that is determined by solving $\partial_\kappa f_1(m,s,\kappa)=0$:
\begin{eqnarray}
    {\kappa}_{1}^{+} &=& 2\frac{1+m}{2}\frac{s}{1+m} \left(1-\frac{s}{1+m}\right)\nonumber \\
    &=& s \left(1-\frac{s}{1+m}\right).
    \label{eq:sp1}
\end{eqnarray}
The first line of \eqref{eq:sp1}, which we have deliberately reported, bears the following direct interpretation. In the thermodynamic limit, a configuration with magnetization $M$ has $\frac{N+M}{2}$ up-spins, which justifies the factor $\frac{1+m}{2}$. The probability that an up-spin is adjacent to a defect is $\frac{s}{1+m}$; thus, the probability that a generic spin is surrounded by a defect is $2 \frac{s}{1+m} \big(1 - \frac{s}{1+m}\big)$, where the factor $2$ accounts for the defect possible being on both sides.

It is worth noting that, from the expressions of ~\eqref{SMeq:conj1}, we can determine the constraint 
\begin{equation}
S\leq \min\left( \frac{N-M}{2},\frac{N+M}{2}-1 \right). 
\end{equation}
In the thermodynamic limit, this translates into $s \leq 1-\vert m \vert$.

\subsubsection{Enumeration of up-spins adjacent to two defects}

We now discuss the enumeration of spins $+1$ adjacent to two defects. The number $\Omega^+_2(N,M,2S,k)$ of configurations of $N$ spins, magnetization $M$, $2S$ defects and $k$ up-spins adjacent to two defects is:
\begin{equation}
        \Omega^+_2(N,M,2S,k) = 
        \left\{\begin{array}{ll}
        \frac{2N}{N-M}\binom{\frac{N-M}{2}}{\frac{N+M}{2}} \quad \text{ if } k=S=\frac{N+M}{2},\\[2mm]
        \frac{2N}{N-M} \binom{\frac{N-M}{2}}{S-k}\binom{\frac{N+M}{2}-S-1}{S-k-1}\binom{\frac{N-M}{2}+k-S}{k}  \text{ otherwise }.
        \end{array}\right.
        \label{eq:conj2}
\end{equation}
As before, $N$ and $M$ must share the same parity. The number of spins $+1$ adjacent to two defects may be either odd or even.

The exponential growth rate of $\Omega^+_2(N,M,2S,k)$, in the large-$N$ limit, is determined by
\begin{equation}
    \lim_{N\rightarrow + \infty} 
    \frac{\log \left( \Omega^+_2(N,M,2S,k) \right)}{N} = \frac{1}{2} f_2(m,s,\kappa),
\end{equation}
where $m, s, \kappa$ are the intensive variables of \eqref{eq:intensive_variables}, and 
\begin{eqnarray}
        f_2(m,s,\kappa) &=&
        (1-m)\log(1-m) - 2(s-2\kappa)\log(s-2\kappa)\nonumber\\
        &+& (1+m-s)\log(1+m-s)\nonumber\\
        &-& (1+m+2\kappa-2s)\log(1+m+2\kappa-2s)\nonumber\\
        &-& 2\kappa \log(2\kappa) - (1-s-m)\log(1-s-m).
\end{eqnarray}
For fixed $m,s$, the function $f_2(m,s,\kappa)$ has a maximum at
\begin{equation}
    {\kappa}_{2}^+ = \frac{1+m}{2}{\left( \frac{s}{1+m}\right)}^{2},
\end{equation}
which is such that $0\leq \kappa_2^+ \leq 1/2$. The lower bound ${\kappa}_{2}^+ = 0$ is attained when $s=0$, corresponding to configurations with a sublinear number of defects. The upper bound ${\kappa}_{2}^+ = 1/2$ is reached when $s=1$ and $m=0$, which corresponds to configurations with a perfect alternation between spins $+1$ and spins $-1$. 

\subsubsection{Enumeration of up-spins adjacent to no defect}

The number $\Omega^+_0(N,M,s,k)$ of configurations of $N$ spins, magnetization $M$, $2S$ defects and $k$ up-spins ($+1$) adjacent to no defect is
\begin{equation}
        \Omega^+_0(N,M,2S,k) = \left\{\begin{array}{ll}
        \frac{2N}{N-M}\binom{\frac{N-M}{2}}{\frac{N+M}{2}} \quad \text{ if } S=\frac{N+M}{2} \text{ and } k=0,\\[2mm]
        \frac{2N}{N-M}\binom{\frac{N-M}{2}}{S} \binom{S}{\frac{N+M}{2}-S-k} \binom{\frac{N+M}{2}-S-1}{k}\text{ otherwise }.
        \end{array}\right.
        \label{eq:conj3}
\end{equation}

In the large-$N$ limit, the asymptotic behavior of $\Omega^+_0(N,M,2S,k)$ is:
    \begin{equation}
    \lim_{N \rightarrow + \infty} \frac{\log\left( \Omega^+_0(N,M,2S,k) \right)}{N} = \frac{1}{2} f_0(m,s,\kappa),
\end{equation}
where 
\begin{equation}
    \begin{split}
    f_0(m,s,\kappa) =  &(1-m)\log(1-m) - (1-m-s)\log(1-m-s)\nonumber\\
    &- 2(1+m-s-2\kappa) \log(1+m-s-2\kappa)\\
    &- (2s+2\kappa -1-m) \log(2s+2\kappa-1-m)\\
    &+(1+m-s)\log(1+m-s) - 2\kappa\log(2\kappa).
    \end{split}
\end{equation}
For fixed $m$ and $s$, the peak of the function $f_0(m,s,\kappa)$ is at
\begin{equation}
   \kappa_{3}^+ = \frac{1+m}{2}{\left( 1-\frac{s}{1+m} \right)}^{2}.
\end{equation}

\subsubsection{Statistics for down-spins}

To derive similar expressions for the enumeration of down-spins ($-1$) adjacent to one, two, or no defects, we note that flipping all the spins in a configuration does not affect the number of defects, but reverses the magnetization. This observation enables us to provide the following statements.

The number $\Omega^-_1(N,M,2S,k)$ of configurations of $N$ spins, magnetization $M$, $2S$ defects and $2k$ down-spins ($-1$) adjacent to a single defect is:
\begin{equation}
    \Omega^-_1(N,M,2S,2k) = \Omega_1^+(N,-M,2S,2k).
\end{equation}

The number $\Omega^-_1(N,M,2S,k)$ of configurations of $N$ spins, magnetization $M$, $2S$ defects and $k$ down-spins ($-1$) adjacent to two defects reads:
\begin{equation}
        \Omega^-_2(N,M,2S,k) = \Omega_2^+(N,-M,2S,k).
\end{equation}

The number $\Omega^-_0(N,M,2S,k)$ of configurations of $N$ spins, magnetization $M$, $2S$ defects and $k$ down-spins ($-1$) adjacent to no defect is:
\begin{equation}
        \Omega^-_0(N,M,2S,k) = \Omega_0^+(N,-M,2S,k).
\end{equation}

\section{Fokker-Planck and Langevin equations}
\label{sec8}

A systematic expansion of the master equation in the large-$N$ limit shows that the Glauber dynamics of the Nagle-Kardar model reduces to a Fokker-Planck equation for the joint probability density $p(m,s;t)$ of the rescaled magnetization $m=M/N$ and fraction of defects $s=2S/N$. The derivation relies on a Taylor expansion of all transition terms in powers of $\epsilon = 1/N$. As detailed in Appendix~\ref{app:FP}, collecting the drift and diffusion contributions that result from the spin-flip channels yields the following two-dimensional Fokker-Planck equation:
\begin{eqnarray}\label{eq:Fokker-Planck}
    &&\frac{\partial p(m,s;t)}{\partial t} = \frac{\partial }{\partial m}\big[D_1(m,s) p(m,s;t)\big]
    + \frac{\partial }{\partial s}[D_2(m,s) p(m,s;t)\big]\nonumber\\
    && \qquad +\epsilon \frac{\partial^2}{\partial m^2} \Big[ D_{11}(m,s) p(m,s;t)\Big]
    +\epsilon \frac{\partial^2}{\partial s^2} \Big[ D_{22}(m,s) p(m,s;t)\Big]\nonumber\\
    && \qquad +2\epsilon \frac{\partial^2}{  \partial m \partial s}\Big[D_{12}(m,s) p(m,s;t)\Big].
\end{eqnarray}
In \eqref{eq:Fokker-Planck}, the drift coefficients are
\begin{eqnarray}
    D_1(m,s) &=& m+c_1 \tanh(x) +  c_2 \tanh(x_{-}) + c_3 \tanh(x_{+})\nonumber\\
    &+& K\epsilon\left( \frac{c_4}{\cosh^2(x)} +  \frac{c_5}{\cosh^2(x_{-})} + \frac{c_6}{\cosh^2(x_{+})}\right),\quad \\
    D_2(m,s) &=& 2s -1 + c_2 \tanh(x_{-})-c_3 \tanh(x_{+})\nonumber\\
    &+& K \epsilon \left( \frac{c_5}{\cosh^2(x_{-})} - \frac{c_6}{\cosh^2(x_{+})} \right),
\end{eqnarray}
while the diffusion coefficients are defined as
\begin{equation}
\begin{split}
& D_{11}(m,s) = 1 - c_4 \tanh(x)
    -c_5 \tanh(x_{-}) - c_6 \tanh(x_{+}),\\
& D_{22}(m,s) = 1 + c_1 -c_5 \tanh(x_{-}) - c_6 \tanh(x_{+}),\\
& D_{12}(m,s) = -m -c_5 \tanh(x_{-}) + c_6\tanh(x_{+}),\\
& D_{21}(m,s) = D_{12}(m,s),
\end{split}
\end{equation}
with $x := h+Km$ and $x_{\pm} := x \pm 2J$ ($T=1$).

Let us now provide the derivation of the Langevin equation corresponding to the Fokker-Planck equation of \eqref{eq:Fokker-Planck}. The derivation is based on Ref.~\cite{risken1989fokker}. We consider a set of $d$ stochastic variables $\{ X_i \}_{i=1}^d$, and the following $d$-dimensional Langevin equation written in Ito convention:
\begin{equation}\label{eq:SM_Langevin_equation}
    \dot{X_i} = h_i \left( \{X_i\}, t \right) + \sqrt{2} \sum_{j=1}^d g_{ij}\left( \{X_i\},t \right)\Gamma_j(t)
\end{equation}
with $i=1,\ldots,d$. In \eqref{eq:SM_Langevin_equation}, $h_i$ and $g_{ij}$ are time-dependent functions of the random variables $X_i$, while $\Gamma_j(t)$ are Gaussian stochastic processes fulfilling the normalizations $\langle \Gamma_i(t)\rangle = 0$ and $\langle \Gamma_j(t) \Gamma_i(t')\rangle = \delta_{ij}\delta(t-t')$. To the Langevin equation (\ref{eq:SM_Langevin_equation}) is associated the Fokker-Planck equation~\cite{risken1989fokker}
\begin{equation}
    \frac{\partial p}{\partial t} = \mathcal{L}_{\rm FP}[p],
\end{equation}
where 
\begin{equation}
\quad \mathcal{L}_{\rm FP} = - \sum_i \frac{\partial}{\partial X_i}D_i\left( \{X_i\}, t\right) + \sum_{i,j} \frac{\partial^2}{\partial X_i \partial X_j} D_{ij}\left( \{X_i\}, t\right)
\end{equation}
and 
\begin{equation}
\begin{split}
& D_i\left( \{X_i\}, t\right) = h_i\left( \{X_i\}, t\right),\\
& D_{ij}\left( \{X_i\}, t\right) = \sum_{k} g_{ik}\left( \{X_i\}, t\right) g_{jk}\left( \{X_i\}, t\right).
\end{split}
\end{equation}
Notice that a given Langevin equation uniquely determines an associated Fokker-Planck equation; however, the reverse is not generally true~\cite{risken1989fokker}.

It follows that the drift terms $h_i( \{X_i\}, t)$ of the Langevin equation corresponding to the Fokker-Planck equation (\ref{eq:Fokker-Planck}) are given by
\begin{equation}
    h_1(m,s) = -D_1(m,s) \quad \text{and} \quad h_2(m,s) = -D_2(m,s)
\end{equation}
with $i,j=1,2$ and $\bm{X}:={(X_1, X_2)}^{\rm tr}={(m,s)}^{\rm tr}$ (`{\rm tr}' denotes transposition). Instead, the diffusion terms $g_{11}, g_{12}, g_{21}, g_{22}$ of \eqref{eq:SM_Langevin_equation} must obey the following relations:
\begin{equation}\label{eq:FP_eq_coefficients}
\begin{split}
& \epsilon D_{11} = g_{11}^2 + g_{12}^{2}, \qquad\quad\,\, \epsilon  D_{22} = g_{22}^2 + g_{21}^{2}, \\
& \epsilon D_{21} = g_{22}g_{12} + g_{21}g_{11}, \quad \epsilon  D_{12} = g_{11}g_{21} + g_{12}g_{22}.
\end{split}
\end{equation}
In \eqref{eq:FP_eq_coefficients}, there are $4$ unknowns but $3$ independent equations. This leads us to set a constraint when choosing the coefficients $g_{ij}$. To do this, we introduce the diffusion matrix
\begin{equation}
    D = \epsilon  \begin{pmatrix}
        D_{11} & D_{12}\\
        D_{21} & D_{22}
    \end{pmatrix},
\end{equation}
and seek for a matrix $G$ such that $D=GG^T$. As there are fewer equations than unknowns, we assume $G$ to be lower triangular:
\begin{equation}
    G = \sqrt{\epsilon} 
    \begin{pmatrix}
        g_{11} & 0\\
        g_{21} & g_{22}
    \end{pmatrix},
\end{equation}
which entails $D_{11} = g_{11}^2$, $D_{22} = g_{21}^2 + g_{22}^2$ and $D_{21} = D_{12} = g_{11} g_{21}$. As a result,
\begin{equation}
    g_{11} = \sqrt{D_{11}}, \quad g_{21} = \frac{D_{21}}{\sqrt{D_{11}}}, \quad g_{22} = \sqrt{ D_{22} - \frac{D_{21}^2}{D_{11}} }.
\end{equation}
It can be checked that $D_{11}>0$ and $D_{22}-D_{21}^2/D_{11}>0$, such that the two-dimensional Langevin equation associated to the Fokker-Planck equation (\ref{eq:Fokker-Planck}) is:
\begin{equation}\label{eq:Langevin}
    \begin{pmatrix}\dot{m} \\ \dot{s}\end{pmatrix} = - 
    \begin{pmatrix}D_{1} \\ D_{2} \end{pmatrix} 
    + \sqrt{2\epsilon} 
    \begin{pmatrix}
    \sqrt{D_{11}} & 0\\
    \frac{D_{12}}{\sqrt{D_{11}}} & \sqrt{D_{22}-\frac{D_{12}^2}{D_{11}}}
\end{pmatrix}
\begin{pmatrix}\Gamma_m \\ \Gamma_s \end{pmatrix}. 
\end{equation}

\section{Bistability \& Metastability}
\label{sec9}

\begin{figure}[t!]
    \centering
    \includegraphics[width=0.99\linewidth]{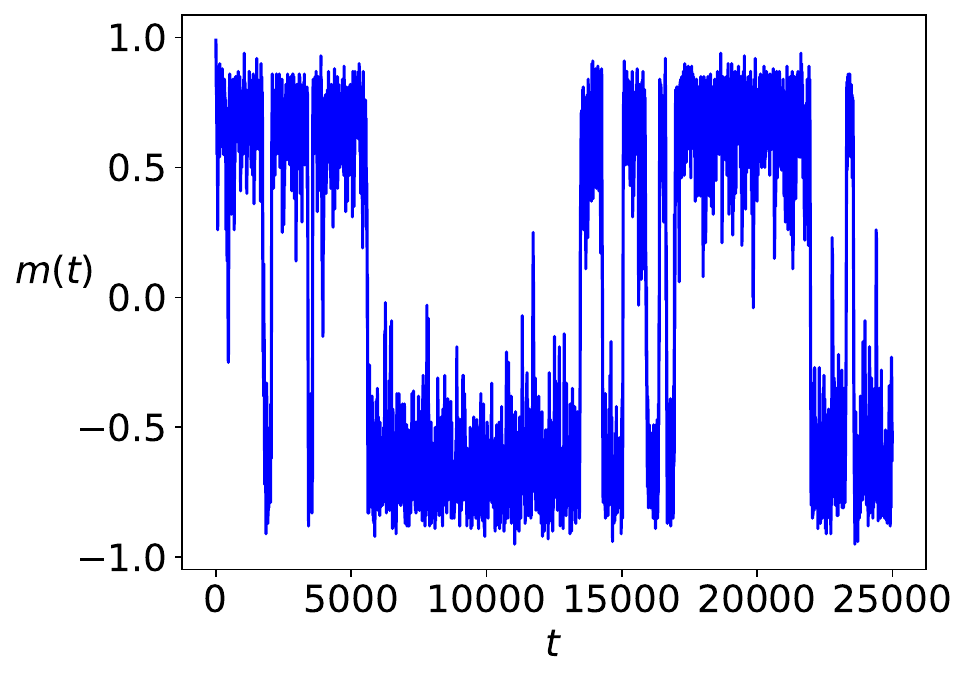}
    \caption{Time-dependence of the magnetization $m(t)$, obtained using the Glauber dynamics for $N=200$. The magnetization shows flips from positive to negative values, at random times overcoming a barrier of $\Delta U = 0.018$. Parameters: $J=K=0.5$ ($T=1$), $h=0$.
    }
    \label{fig:m_bistability}
\end{figure}

As we have seen in Section~\ref{sec4}, the quasi-potential associated to the Nagle-Kardar model in the large-$N$ limit can display two local minima separated by a local maximum or saddle-point, see Figs.~\ref{fig:2D_potential} and \ref{figSM:Large_deviation_magnetization}. When started in a local minimum, at all finite temperatures, the system can overcome the barrier that separates the two minima and fall into the other minimum in a finite time. The heights of the two minima depend on parameters, and one finds both bistable situations where the minima have the same height (e.g., $h=0$) and metastable ones where one of the two minima is higher. The first-passage time $\tau$ from one minimum to the other is a random variable and its derivation for $N$-body systems is still an open question. However, using both analytical approximations and numerical simulations, it has been found that, for systems with long-range interactions, including the Nagle-Kardar model, its average $\langle\tau\rangle$ obeys the Arrhenius law~\cite{griffiths1966relaxation,antoni2004first,chavanis2005lifetime,mukamel2005breaking,mori2010asymptotic,Saadat2023,chavanis2026thermal}
\begin{eqnarray}
    \langle \tau \rangle \sim e^{N \Delta U},
\end{eqnarray}
where $\Delta U$ is the height of the barrier separating the two minima of $U(m,s)$ [\eqref{eq:potentialU}]. Under both microscopic Glauber dynamics and macroscopic Langevin dynamics, the magnetization evolves in time, as shown in Fig.~\ref{fig:m_bistability} for bistable states. By collecting several trajectories of this kind, we have evaluated the average first-passage time in both the bistable and metastable case, with the aim of showing that averages obtained along trajectories of the Langevin equations agree with those obtained from microscopic dynamics.

\begin{figure}[t!]
    \centering
    \includegraphics[width=0.85\linewidth]{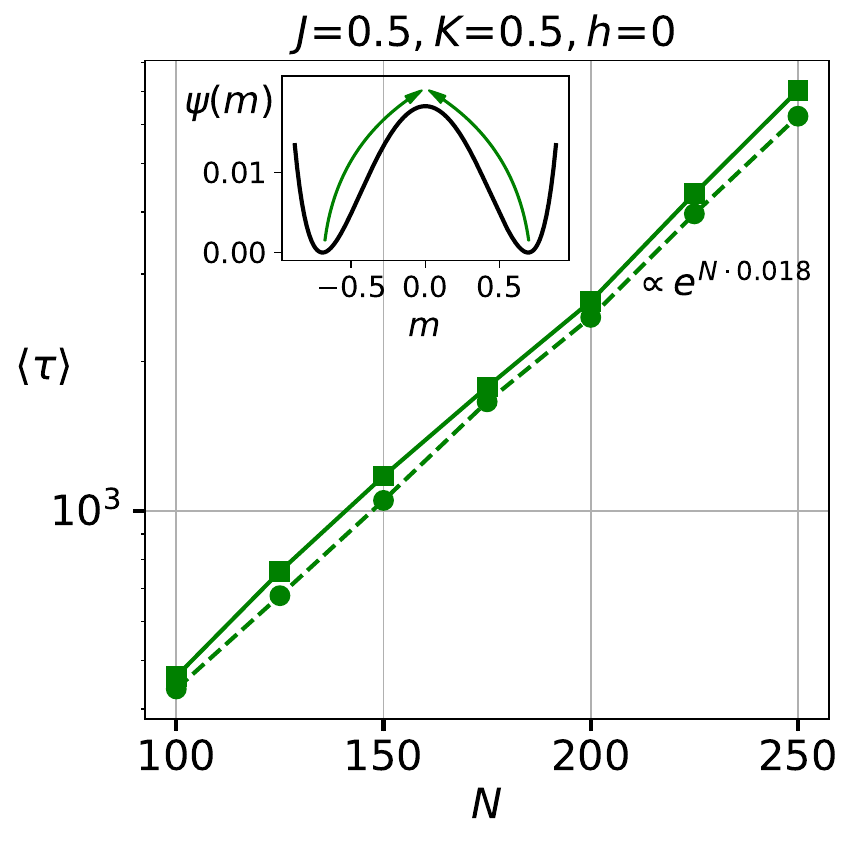}
    \includegraphics[width=0.85\linewidth]{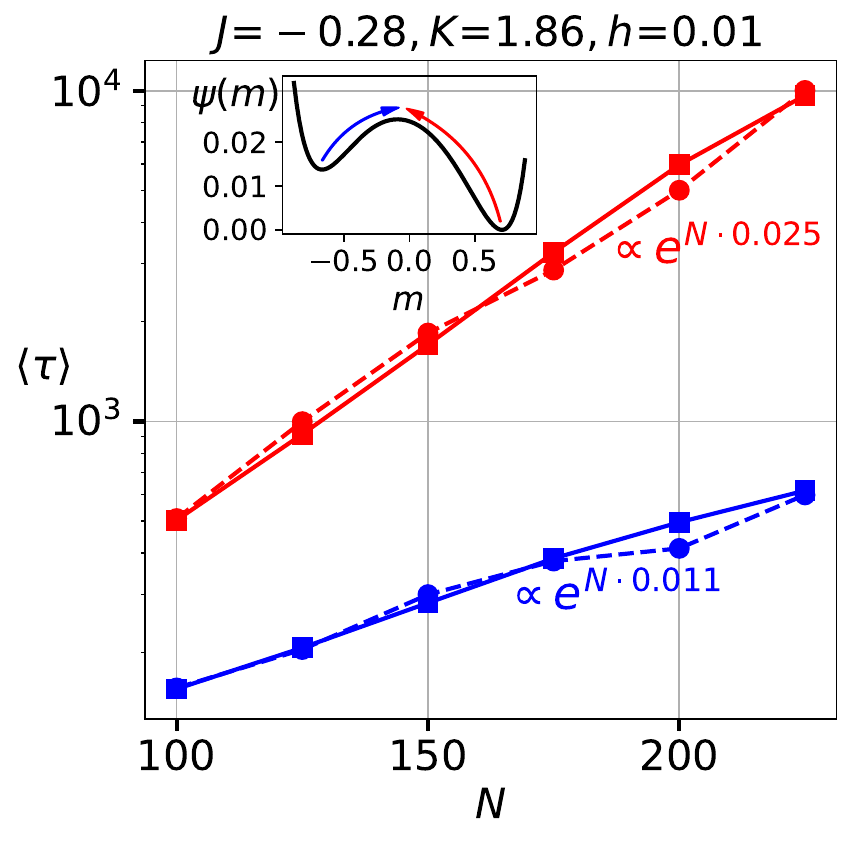}
    \caption{
    (Top) \textit{Bistability}: Average first passage time $\langle\tau\rangle$ versus $N$, for $h=0$, $J=0.5$, $K=0.5$ ($T=1$) in the Nagle-Kardar model. We compare the Glauber (green squares) and Langevin (green dots) dynamics, obtaining a very good agreement. The plot is in semi-log scale to highlight the exponential growth with $N$, and the lines are drawn to guide the eyes.
    (Bottom) \textit{Metastability}: We repeat the same analysis with non-zero $h$ to study metastability, with $h=0.01$, $J=-0.28$, $K=1.86$. Inset of both panels: Plot of the large deviation function $\psi(m)$ [\eqref{eq:LDVmagn}], and transitions between the minima of the quasi-potential. 
    Typically we average over tens of trajectories for a physical time of order $10^4$.
    }
    \label{fig:Arrhenius}
\end{figure}

Fig.~\ref{fig:Arrhenius} displays $\langle\tau\rangle$ versus $N$. We compare Glauber with Langevin dynamics in the bistable case, with zero external field $h=0$ (top panel), and in the metastable one with $h\neq 0$ (bottom panel).
To depict the transitions between the two minima of the quasi-potential $U(m,s)$ (green arrows for the bistable case and blue/red arrows for the metastable one), we report the large deviation function $\psi(m)$ [\eqref{eq:LDVmagn}] in the inset of both panels of Fig.~\ref{fig:Arrhenius}. In particular, in the inset of the top panel, due to the symmetry of $\psi(m)$, the left-right passage time is equal to the right-left one. Using \eqref{eq:potentialU}, the exponential growth rate is determined by the variation $\Delta U = 0.018$, calculated between the stable and the saddle points in the $(m,s)$ plane. On the other hand, in the inset of the bottom panel, $\psi(m)$ is asymmetric and we need to compute two different scalings for the left-right (blue lines) and right-left (red lines) passage times. In such case, the rates of exponential increase are equal to $\Delta U=0.011$ (blue) and $\Delta U=0.025$ (red), respectively.

In both cases (bistability and metastability), we can confirm that the Glauber dynamics (squares) gives the same result as the Langevin dynamics (dots). The calculation of the prefactor in front of $e^{N \Delta U}$ would need a deeper analysis of both the drift and diffusion terms of the Langevin equation~\cite{lee2022non}.

\section{Conclusions}

In a companion Letter~\cite{deKemmeterLetter2025}, we have shown that the critical dynamics of the Nagle-Kardar model belongs to the universality class of diffusive non-conserved order parameter dynamics~\cite{hohenberg1977theory}. In order to obtain these results, we have used a coarse-graining procedure that starts with a microscopic Glauber dynamics and reaches a Fokker-Planck equation for the magnetization and density of defects.

In this paper, we have reported all the details of the derivation of the Fokker-Planck equation [\eqref{eq:Fokker-Planck}] and of the related Langevin equation [\eqref{eq:Langevin}]. This derivation is based on the crucial assumption that the dynamics visits microstates on a faster time scale compared with the slow relaxation of macroscopic variables. From the technical point of view, the derivation also requires the solution of some combinatorial problems implied by the counting of microscopic configurations, cfr.~Section \ref{sec7}. Finally, in order to complete the study of the equilibrium canonical solution of the Nagle-Kardar model, we have computed the finite-$N$ expression for fluctuations of macroscopic variables and of the correlation function. We have also added a section (Section \ref{sec9}) on the bistable and metastable behaviors of the model when a free-energy barrier is present.

\subsection*{Acknowledgments}

The authors would like to thank David Mukamel for carefully reading the manuscript and providing us with helpful comments. We acknowledge financial support from the PRIN 2022 PNRR P2022XPT32 ``Regulation of striated muscle: a research bridging single molecule to organ (ReStriMus)'' funded by the European Union---Next Generation EU, and from the MUR PRIN2022 project BECQuMB Grant No.~20222BHC9Z.

\appendix

\section{Exact expression of $P_N(M)$ for finite-size systems}
\label{app:JPDF}

The probability distribution $P_N(M)$ for the magnetization $M$ can be expressed in terms of hypergeometric functions. This can be seen by computing the summation
\begin{eqnarray}
    \sum_{S=1}^{\lfloor N/2 \rfloor}
    \binom{\frac{N+M}{2}-1}{S-1} \binom{\frac{N-M}{2}}{S}
    \frac{e^{\frac{J}{T}(N-4S) + \frac{K}{2NT}M^2 + \frac{h}{T}M }}{Z_N}\nonumber\\
    = \frac{e^{ \frac{J}{T}N + \frac{K}{2NT}M^2 + \frac{h}{T} M }}{Z_N} \sum_{S=1}^{\lfloor N/2 \rfloor}
    \binom{\frac{N+M}{2}-1}{S-1} \binom{\frac{N-M}{2}}{S} x^S,
\end{eqnarray}
with $x=e^{-4J/T}$. Let us first assume $N$ to be even. In that case, we have:
\begin{eqnarray}
    &&\sum_{S=1}^{N/2}\binom{\frac{N+M}{2}-1}{S-1} \binom{\frac{N-M}{2}}{S} x^S 
    = x \sum_{S=0}^{N/2-1}\binom{\frac{N+M}{2}-1}{S} \binom{\frac{N-M}{2}}{S+1} x^S \nonumber\\
    &&= x \sum_{S=0}^{N/2-1} \frac{\big(\frac{N+M}{2}-1\big)!}{S! \big(\frac{N+M}{2}-S-1\big)!} \frac{\big(\frac{N-M}{2}\big)!}{(S+1)! \big(\frac{N-M}{2}-S-1\big)!} x^S \nonumber \\
    &&= x \frac{N-M}{2}\sum_{S=0}^{N/2-1} \frac{
    {\big(\frac{N+M}{2}-1\big)}_{\underline{S}}}{S! } \frac{{\big(\frac{N-M}{2}-1\big)}_{\underline{S}}}{ {(2)}_{\overline{s}}} x^{S},\quad
\end{eqnarray}
where 
\begin{eqnarray}
    (\alpha)_{\underline{S}} &:=& \alpha (\alpha-1)\cdots(\alpha-S+1) = \frac{\alpha!}{(\alpha - S)!},\\
    (\alpha)_{\overline{S}} &:=& \alpha (\alpha+1) \cdots(\alpha+S-1) = \frac{(\alpha + S -1)!}{(\alpha - 1)!}
\end{eqnarray}
are respectively the falling and the rising factorials (the rising factorial is also called the Pochhammer symbol), each given by the product of $S$ terms. The rising and falling factorials are related by the following identity:
\begin{equation}
    (\alpha)_{\underline{S}} = {(-1)}^S (-\alpha)_{\overline{S}}.
\end{equation}
Using this identity, it holds that
\begin{eqnarray}
    &&\sum_{S=1}^{N/2}\binom{\frac{N+M}{2}-1}{S-1} \binom{\frac{N-M}{2}}{S} x^S \nonumber\\
    &&= x \frac{N-M}{2}\sum_{S=0}^{N/2-1}\frac{
    {\big(1-\frac{N+M}{2}\big)}_{\overline{S}}{\big(1-\frac{N-M}{2}\big)}_{\overline{S}}
    }{{(2)}_{\overline{S}}} \frac{ x^S }{S!}.
\end{eqnarray}
One can extend the summation to infinity without modifying the result, which leads to
\begin{eqnarray}
    &&\sum_{S=1}^{N/2}\binom{\frac{N+M}{2}-1}{S-1} \binom{\frac{N-M}{2}}{S} x^s \nonumber\\
    &&= x \frac{N-M}{2} \sum_{S=0}^\infty \frac{
    {\big(1-\frac{N+M}{2}\big)}_{\overline{S}}{\big(1-\frac{N-M}{2}\big)}_{\overline{S}}
    }{{(2)}_{\overline{S}}} \frac{ x^S }{S!}\nonumber\\
    &&=x\frac{N-M}{2}\,
    \setlength\arraycolsep{1pt}
{}_2F_1\Big(1-\frac{N+M}{2},1-\frac{N-M}{2};2;x\Big),
\end{eqnarray}
where ${}_2F_1(a,b,c;\zeta)$ denotes the hypergeometric function defined by the series \cite{abramowitz1965handbook}
\begin{equation}
    {}_2F_1(a,b;c;\zeta) := \sum_{n=0}^\infty \frac{(a)_{\overline{n}} (b)_{\overline{n}}}{(c)_{\overline{n}}}\frac{\zeta^n}{n!}.
\end{equation}
The same expression is obtained when $N$ is odd. We now turn our attention to the second sum given by
\begin{equation}
    \frac{e^{\frac{J}{T} N + \frac{K}{2NT}M^2 + \frac{h}{T}M }}{Z} \sum_{S=1}^{\lfloor N/2 \rfloor}
    \binom{\frac{N-M}{2}-1}{S-1} \binom{\frac{N+M}{2}}{S}
    x^{S}.
\end{equation}
Following the same reasoning as above, we get:
\begin{eqnarray}
&&\sum_{S=1}^{\lfloor N/2 \rfloor}
\binom{\frac{N-M}{2}-1}{S-1} \binom{\frac{N+M}{2}}{S}x^S = x \sum_{S=0}^{\lfloor N/2 \rfloor -1} \binom{\frac{N-M}{2}-1}{S} \binom{\frac{N+M}{2}}{S+1}x^S \nonumber \\
&& = x \frac{N+M}{2} \sum_{S=0}^{\lfloor N/2 \rfloor -1} \frac{
{\big(\frac{N-M}{2}-1\big)}_{\underline{S}} {\big(\frac{N+M}{2}-1\big)}_{\underline{S}}}{{(2)}_{\overline{S}} }\frac{x^S}{S!} \nonumber\\
&& = x \frac{N+M}{2} \setlength\arraycolsep{1pt}
{}_2F_1\Big(1-\frac{N+M}{2},1-\frac{N-M}{2};2;x\Big).
\end{eqnarray}
Combining all the computed terms, we finally obtain:
\begin{eqnarray}\label{SMeq:Pnmhyp}
   P_N(M) &=& N \frac{e^{ \frac{J}{T} (N-4) + \frac{K}{2NT}M^2 + \frac{h}{T} M }}{Z_N}\nonumber\\
   &\times& \setlength\arraycolsep{1pt}
   {}_2F_1\left( 1-\frac{N+M}{2},1-\frac{N-M}{2},2;e^{-4 J/T} \right).\quad
\end{eqnarray}

\section{Asymptotic expressions of $P_N(M)$ in the large-$N$ limit}\label{app:LD}

Here, we provide an asymptotic expression for the probability distribution $P_N(M)$ in the large-$N$ limit. This is done by using the following asymptotic expression of the hypergeometric functions derived in Ref.~\cite{cvitkovic2017asymptotic}:
\begin{equation}\label{eqSM:asymptotic_hypergeom}
{}_2F_1\left( a - \varepsilon \lambda,\ b - \lambda; c; \zeta \right) 
\sim \frac{\Gamma(c) (\varepsilon \lambda)^{\frac{1}{2} - c}}{\sqrt{2\pi |\varepsilon - 1| \tilde{\sigma}(\zeta)}} 
\frac{(1 - \xi)^{c - a + \varepsilon \lambda}}{(-\xi)^{\varepsilon \lambda - a}(1 - \zeta\xi)^{b - \lambda}}
\end{equation}
that is valid for $a,b,c$ real and $\lambda \to \infty$. In \eqref{eqSM:asymptotic_hypergeom},
\begin{equation}
    \xi = \frac{1-\epsilon}{2} - \frac{\vert 1-\epsilon \vert}{2} \tilde{\sigma}(\zeta)
\end{equation}
with
\begin{equation}
\tilde{\sigma}(\zeta)= \sqrt{1+\frac{4\epsilon}{{(\epsilon-1)}^{2}\zeta}}
\end{equation}
In our case, $\lambda := \lambda(m)=\frac{N-M}{2}=N\frac{1-m}{2}$, $\epsilon := \epsilon(m) =\frac{N+M}{N-M}=\frac{1+m}{1-m}$, $a=1$, $b=1$, $c=2$, $\zeta=e^{-4J/T}$. Hence, inserting the latter parameters into \eqref{SMeq:Pnmhyp} leads, after a few simplifications, to $P_N(M) \asymp e^{-N \psi(m)}$, with $\psi(m)$ given by \eqref{eq:LDVmagn} in the main text.

\section{Asymptotic of the pair-correlation function}\label{app:gr}

In this Appendix, we extract the asymptotic behavior of the pair-correlation function $g(r)=\langle s_i s_r \rangle - \langle s_i \rangle \langle s_r \rangle$ along the critical line, in the $h=0$ plane. We set $T=1$ without loss of generality. Since $h=0$, we have $\langle s_i \rangle=0$, from which we obtain:
\begin{equation}
    g(r) = \langle s_i s_{i+r}\rangle = \frac{ I_1(N) }{ I_3(N) } + \frac{ I_2(N) }{ I_3(N) },
\end{equation}
with 
\begin{equation}
    \begin{split}
        I_1(N) &= \int_\mathbb{R} \sin^{2}\Big( 2\phi(x) \Big)  \left( \frac{\lambda_{-}(x)}{\lambda_{+}(x)} \right)^r e^{NF(x)} dx,\\
        I_2(N) &= \int_\mathbb{R} \cos^{2}\Big( 2\phi(x) \Big) e^{NF(x)} dx,\\
        I_3(N) &= \int_\mathbb{R} e^{NF(x)} dx
    \end{split}
\end{equation}
and
\begin{equation}
    \lambda_\pm(x) = K^{-1/2} \left( \cosh(Kx) \pm \sqrt{K^2+\sinh^2 (Kx)} \right),
\end{equation}
where $\cot\Big( 2\phi(x) \Big) = e^{2J} \sinh(Kx)$.
Along the critical line $e^{-2J} = K$, we have:
\begin{equation}
    \begin{split}
        &\sin^{2}\Big( 2\phi(x) \Big) = \frac{K^2}{K^2 + \sinh^2(Kx)}, \\
        &\cos^{2}\Big( 2\phi(x) \Big) = \frac{\sinh^2(Kx) }{K^2 + \sinh^2(Kx)}.
    \end{split}
\end{equation}
To obtain the asymptotic behavior of $g(r)$ at large $N$, we determine the behavior of $F(x) = - \frac{K}{2 T} x^2 +\log\left( \lambda_{+}(x) \right)$ around the saddle-point $x^*=0$:
\begin{eqnarray}
    F(x) &=& \log \left( \frac{1+K}{\sqrt{K}} \right)
    + \frac{K(K^2-3)}{24}\nonumber\\
    &+& x^4\frac{K(K^4-30K^2+45)}{720}x^6 + O(x^8).
\end{eqnarray}
Moreover, at $x^*=0$, one has:
\begin{equation}
    \cos^{2}\Big( 2\phi(x) \Big) = x^2 + \left( \frac{K^2}{3}-1 \right)x^4 
    + \left( \frac{2K^4}{45}-\frac{2K^2}{3} + 1 \right)x^6 + O(x^8),
\end{equation}
and
\begin{equation}
    \sin^{2}\Big( 2\phi(x) \Big) \left( \frac{\lambda_{-}(x)}{\lambda_{+}(x)} \right)^r = \left( \frac{1-K}{1+K} \right)^{r} \left( 1-(1+rK)x^2 + O(x^4) \right).
\end{equation}

At this point, with the aim to extract the asymptotic of the correlation function, we use the following result, with $F(x)=F(0)-cx^{2m} + O(x^{2m+2})$, $c>0$ and $g(x)=g_0x^{q} + O(x^{q+1})$, $q\ge 0$:
\begin{equation}\label{eq:integral_kernel_Fx}
    \int_\mathbb{R} g(x) e^{NF(x)} dx \sim e^{NF(0)} \frac{g_0}{2m} \Gamma\left( \frac{q+1}{2m} \right) {(Nc)}^{-\frac{q+1}{2m}}.
\end{equation}
Accordingly, by means of \eqref{eq:integral_kernel_Fx}, for $0<K<\sqrt{3}$ we find:
\begin{equation}
    \begin{split}
    I_1(N) & \sim \left( \frac{1-K}{1+K} \right)^r \frac{\Gamma(1/4)}{4} \left( \frac{K(3-K^2)}{24}\right)^{-1/4} N^{-1/4} e^{NF(0)},\\
    I_2(N) & \sim \frac{\Gamma(3/4)}{4} \left( \frac{K(3-K^2)}{24}\right)^{-3/4} N^{-3/4} e^{NF(0)},\\
    I_3(N) & \sim \frac{\Gamma(1/4)}{4} \left( \frac{K(3-K^2)}{24}\right)^{-1/4} N^{-1/4} e^{NF(0)}.
    \end{split}
\end{equation}
As a result, the following asymptotic behavior is obtained:
\begin{equation}\label{eq:appasymg}
    g(r) \sim \left( \frac{1-K}{1+K} \right)^r + \frac{\Gamma(3/4)}{\Gamma(1/4)} \left( \frac{24}{K(3-K^2)}\right)^{1/2} \frac{1}{\sqrt{N}}.
\end{equation}

At the tricritical point ($K=\sqrt{3}$), instead, one has:
\begin{equation}
    \begin{split}
    I_1(N) & \sim \left( \sqrt{3} - 2 \right)^r\,\frac{\Gamma(1/6)}{6} \left( \frac{\sqrt{3}}{20} \right)^{-1/6} N^{-1/6} e^{NF(0)},\\
    I_2(N) & \sim \frac{\Gamma(1/2)}{6} \left( \frac{\sqrt{3}}{20} \right)^{-1/2} N^{-1/2} e^{NF(0)},\\
    I_3(N) & \sim \frac{\Gamma(1/6)}{6} \left( \frac{\sqrt{3}}{20}\right)^{-1/6} N^{-1/6} e^{NF(0)},
    \end{split}
\end{equation}
which leads to the following asymptotic behavior at the tricritical point:
\begin{equation}\label{eq:appasympgTP}
    g(r) \sim \left( \sqrt{3}-2 \right)^r + \frac{\Gamma(1/2)}{\Gamma(1/6)} \left( \frac{\sqrt{3}}{20} \right)^{-1/3} N^{-1/3}.
\end{equation}

\onecolumngrid
\section{Derivation of the Fokker-Planck equation from the Master equation}\label{app:FP}

In this section, we derive the Fokker-Planck equation associated to the Glauber dynamics of the Nagle-Kardar model. To this end, we consider the limit of large $N$ and perform a Taylor expansion of the master equation (\ref{SMeq:dPNdt2}) in terms of the rescaled variables $m=M/N$ and $s=2S/N$. Let us start with the term $A_{M+2\rightarrow M}^{2S\rightarrow 2S}  B_1^{+}(N,M+2,2S) P_N(M+2,2S;t) - A_{M\rightarrow M-2}^{2S\rightarrow 2S} B_1^{+}(N,M,2S) P_N(M,2S;t)$; for all the other terms of the master equation the procedure is the same. We thus use the following Taylor expansions:
\begin{eqnarray}
   P_N(M+2,2S;t) &=& p(m,s;t) + \partial_m p(m,s;t) \Big(\frac{2}{N}\Big) + \frac{1}{2} \frac{\partial^2}{\partial m^2}p(m,s;t) {\Big(\frac{2}{N}\Big)}^2 + O\left( \frac{1}{N^3}\right),\\
   B_1^+(N,M+2,2S) &=& \kappa_1^+(m,s) + \partial_m \kappa_1^+(m,s) \Big(\frac{2}{N}\Big) + \frac{1}{2} \frac{\partial^2}{\partial m^2}\kappa_1^+(m,s) {\Big(\frac{2}{N}\Big)}^2 + O\left( \frac{1}{N^3}\right),\\
   A_{M+2\rightarrow M}^{2S\rightarrow 2S} &=& \frac{
   e^{ -[K (M+1)/N + h]/T }
   }{
   2\cosh \{ [K(M+1)/N + h]/T \}
   } = f_1^{+}(m) + \frac{\partial f_1^{+}(m)}{\partial m}\frac{1}{N} + \frac{1}{2}\frac{\partial^2 f_1^{+}(m)}{\partial{m^2}} {\Big(\frac{1}{N}\Big)}^2 + O\left( \frac{1}{N^3}\right),\\
   A_{M\rightarrow M-2}^{2S\rightarrow 2S} &=& \frac{
   e^{ -[K (M-1)/N + h]/T }
   }{
   2\cosh \{ [K(M-1)/N + h]/T \}
   } = f_1^{+}(m) - \frac{\partial f_1^{+}(m)}{\partial m}\frac{1}{N} + \frac{1}{2}\frac{\partial^2 f_1^{+}(m)}{\partial{m^2}} {\Big(\frac{1}{N}\Big)}^2 + O\left( \frac{1}{N^3}\right),
\end{eqnarray}
where
\begin{equation}
    f_1^{\pm}(m) = \frac{
    e^{ \mp (K m + h)/T }
    }{
    2\cosh [ (K m + h)/T ]
    }.
\end{equation}
Setting $\epsilon=1/N$, we thus obtain (the dependence on $m$ is omitted unless necessary):
\begin{eqnarray}\label{eq:calculation_FP_1}
    && A_{M+2\rightarrow M}^{2S\rightarrow 2S}  B_1^{+}(N,M+2,2S) P_N(M+2,2S;t) = \kappa_1^+ f_1^+ p+ \epsilon \left[ 2 \kappa_1^+ f_1^+ \frac{\partial p}{\partial m} + \kappa_1^+ \frac{\partial f_1^+}{\partial m} p + 2 \frac{\partial \kappa_1^+}{\partial m} f_1^+ p\right]\nonumber\\ 
    && +\epsilon^2 \left[ 2 \kappa_1^+ f_1^+ \frac{\partial^2 p}{\partial m^2} + 2 \kappa_1^+ \frac{\partial f_1^+}{\partial m} \frac{\partial p}{\partial m} + \frac{1}{2} \kappa_1^+ \frac{\partial^2 f_1^+}{\partial m^2} p + 4 \frac{\partial \kappa_1^+}{\partial m} f_1^+ \frac{\partial p}{\partial m} + 2 \frac{\partial \kappa_1^+}{\partial m} \frac{\partial f_1^+}{\partial m} p + 2 \frac{\partial^2 \kappa_1^+}{\partial m^2}f_1^+ p \right],
\end{eqnarray}
and
\begin{equation}\label{eq:calculation_FP_2}
    A_{M\rightarrow M-2}^{2S\rightarrow 2S}  B_1^{+}(N,M,2S) P_N(M,2S;t) = \kappa_1^+\left[ f_1^+ - \frac{\partial f_1^+}{\partial m} \epsilon + \frac{\epsilon^2}{2} \frac{\partial^2f_1^+}{\partial m^2} \right]p.
\end{equation}
From Eqs.~(\ref{eq:calculation_FP_1})-(\ref{eq:calculation_FP_2}) we deduce that
\begin{eqnarray}
    &&A_{M+2\rightarrow M}^{2S\rightarrow 2S}  B_1^{+}(N,M+2,2S) P_N(M+2,2S;t) - A_{M\rightarrow M-2}^{2S\rightarrow 2S}  B_1^{+}(N,M,2S) P_N(M,2S;t)=\nonumber\\ 
    &&= 2 \epsilon \frac{\partial }{\partial m}\big(\kappa_1^+ f_1^+ p\big) + 2\epsilon^2 f_1^+ \frac{\partial^2}{\partial m^2} \big(\kappa_1^+ p\big) + 2\epsilon^2 \frac{\partial f_1^+}{\partial m} \frac{\partial}{\partial m}\big(\kappa_1^+ p\big)\nonumber\\
    &&=2\epsilon \frac{\partial }{\partial m} \big(\kappa_1^+ f_1^+ p\big) + 2\epsilon^2 \frac{\partial}{\partial m} \Big[f_1^+ \frac{\partial }{\partial m} \big(\kappa_1^+ p \big)\Big].
\end{eqnarray}
Similarly, by symmetry, we have:
\begin{eqnarray}
    && A_{M-2\rightarrow M}^{2S\rightarrow 2S} 
    B_1^{-}(N,M-2,2S) P_N(M-2,2S;t)-A_{M\rightarrow M+2}^{2S\rightarrow 2S} 
    B_1^{-}(N,M,2S)P_N(M,2S;t)=\nonumber\\
    && = -2\epsilon \frac{\partial (\kappa_1^- f_1^{-} p )}{\partial m} + 2\epsilon^2 \frac{\partial }{\partial m} \Big[f_1^{-} \frac{\partial}{\partial m}(\kappa_1^- p )\Big].
\end{eqnarray}
Using the procedure above, we also determine that 
\begin{eqnarray}
    && A_{M\pm 2 \rightarrow M}^{2S+2\rightarrow 2S}  B_2^{\pm}(N,M\pm 2,2S+2) P_N(M\pm2,2S+2;t) - A_{M\rightarrow M \mp 2}^{2S \rightarrow 2S-2} B_2^{\pm}(N,M,2S) P_N(M,2S;t)= \nonumber\\
    &&= \pm 2\epsilon \frac{\partial (f_2^{\pm} \kappa_2^\pm p) }{\partial m} 
    + 2\epsilon \frac{\partial (f_2^\pm \kappa_2^\pm p)}{\partial s} +2\epsilon^2 \frac{\partial}{\partial m} \left[ f_2^\pm\frac{\partial (\kappa_2^\pm p)}{\partial m}\right]
    + 2\epsilon^2 f_2^\pm \frac{\partial^2 (\kappa_2^\pm p)}{\partial s^2} 
    \pm 2\epsilon^2 \frac{\partial f_2^\pm}{\partial m} \frac{\partial (\kappa_2^\pm p)}{\partial s}
    \pm 4\epsilon^2 f_2^\pm \frac{\partial^2 (\kappa_2^\pm p)}{\partial m \partial s}
\end{eqnarray}
with the functions $f_2^\pm (m)$ given by
\begin{equation}
    f_2^{\pm}(m) = \frac{
    e^{ \mp (K m + h \mp 2J)/T }
    }{
    2\cosh [ (K m + h \mp 2J)/T ]
    },
\end{equation}
and
\begin{eqnarray}
    &&A^{2S-2 \rightarrow 2S}_{M\pm2 \rightarrow M} B_0^{\pm}(N,M\pm 2,2S - 2) P_N(M\pm2,2S-2;t) - A_{M\rightarrow M \mp 2}^{2S \rightarrow 2S+2} B_0^{\pm}(N,M,2S) P_N(M,2S;t)= \nonumber\\
    && = \pm 2 \epsilon \frac{\partial (\kappa_0^\pm f_0^\pm p) }{\partial m} 
    - 2\epsilon \frac{\partial (\kappa_0^\pm f_0^\pm p)}{\partial s} 
    + 2\epsilon^2 \frac{\partial}{\partial m} \left[ f_0^\pm \frac{\partial (\kappa_0^\pm p)}{\partial m}\right] 
    \mp 2\epsilon^2 \frac{\partial f_0^+}{\partial m} \frac{\partial (\kappa_0^\pm p)}{\partial s}
    + 2 \epsilon^2 f_0^\pm \frac{\partial^2 (\kappa_0^\pm p)}{\partial s^2} \mp 4\epsilon^2 f_0^\pm \frac{\partial^2 (\kappa_0^\pm p)}{\partial s \partial m},
\end{eqnarray}
where 
\begin{equation}
    f_0^{\pm}(m) = \frac{
    e^{ \mp (K m + h \pm 2J)/T }
    }{
    2\cosh [ (K m + h \pm 2J)/T ]
    }.
\end{equation}

In conclusion, gathering all the terms together, we obtain the Fokker-Planck equation for the Nagle-Kardar model:
\begin{eqnarray}
        \frac{\partial}{\partial t}p(m,s;t) &=& \frac{\partial }{\partial m}\big[D_1(m,s) p(m,s;t)\big]
        + \frac{\partial}{\partial s}[D_2(m,s) p(m,s;t)\big]\nonumber\\
        &+& \epsilon \frac{\partial^2}{\partial m^2} \Big[ D_{11}(m,s) p(m,s;t)\Big]
        +\epsilon \frac{\partial^2}{\partial s^2} \Big[ D_{22}(m,s) p(m,s;t)\Big] +2\epsilon \frac{\partial^2}{  \partial m \partial s}\Big[D_{12}(m,s) p(m,s;t)\Big],
        \label{eqSM:FKP}
\end{eqnarray}
where the drift terms are equal to
\begin{eqnarray}
    D_1(m,s) &=& m+c_1 \tanh(x) +  c_2 \tanh(x_{-}) + c_3 \tanh(x_{+}) + K\epsilon\left( \frac{c_4}{\cosh^2(x)} +  \frac{c_5}{\cosh^2(x_{-})} + \frac{c_6}{\cosh^2(x_{+})}\right),\nonumber\\
    D_2(m,s) &=& 2s -1 + c_2 \tanh(x_{-})-c_3 \tanh(x_{+})+ K \epsilon \left( \frac{c_5}{\cosh^2(x_{-})} - \frac{c_6}{\cosh^2(x_{+})} \right),
\end{eqnarray}
while the diffusion terms are defined by
\begin{eqnarray}
&& D_{11}(m,s) = 1 - c_4 \tanh(x)
    -c_5 \tanh(x_{-}) - c_6 \tanh(x_{+}),\nonumber\\
&& D_{22}(m,s) = 1 + c_1 -c_5 \tanh(x_{-}) -c_6 \tanh(x_{+}),\nonumber\\
&& D_{12}(m,s) = -m -c_5 \tanh(x_{-}) + c_6\tanh(x_{+})
\end{eqnarray}
with $x := h+Km$ and $x_{\pm} := x \pm 2J$. Finally, the coefficients $c_i$ ($i=1,\ldots,6$) entering the drift and diffusion terms have the following expressions: 
\begin{align}
& c_1 = -2s\left(1 + \frac{s}{m^2 - 1}\right), \qquad c_2 = \frac{m - 1}{2} + s + \frac{s^2}{m^2 - 1}, \qquad\quad c_3 = -\frac{m + 1}{2} + s + \frac{s^2}{m^2 - 1},\\
& c_4 = -2 \frac{ms^2}{m^2 - 1}, \qquad\qquad\quad c_5 = \frac{m - 1}{2} + s + \frac{m s^2}{m^2 - 1}, \qquad\quad c_6 = \frac{m + 1}{2} - s + \frac{m s^2}{m^2 - 1}.
\end{align}

\twocolumngrid 
\bibliography{bibliography}

\begin{thebibliography}{41}%
\makeatletter
\providecommand \@ifxundefined [1]{%
 \@ifx{#1\undefined}
}%
\providecommand \@ifnum [1]{%
 \ifnum #1\expandafter \@firstoftwo
 \else \expandafter \@secondoftwo
 \fi
}%
\providecommand \@ifx [1]{%
 \ifx #1\expandafter \@firstoftwo
 \else \expandafter \@secondoftwo
 \fi
}%
\providecommand \natexlab [1]{#1}%
\providecommand \enquote  [1]{``#1''}%
\providecommand \bibnamefont  [1]{#1}%
\providecommand \bibfnamefont [1]{#1}%
\providecommand \citenamefont [1]{#1}%
\providecommand \href@noop [0]{\@secondoftwo}%
\providecommand \href [0]{\begingroup \@sanitize@url \@href}%
\providecommand \@href[1]{\@@startlink{#1}\@@href}%
\providecommand \@@href[1]{\endgroup#1\@@endlink}%
\providecommand \@sanitize@url [0]{\catcode `\\12\catcode `\$12\catcode `\&12\catcode `\#12\catcode `\^12\catcode `\_12\catcode `\%12\relax}%
\providecommand \@@startlink[1]{}%
\providecommand \@@endlink[0]{}%
\providecommand \url  [0]{\begingroup\@sanitize@url \@url }%
\providecommand \@url [1]{\endgroup\@href {#1}{\urlprefix }}%
\providecommand \urlprefix  [0]{URL }%
\providecommand \Eprint [0]{\href }%
\providecommand \doibase [0]{https://doi.org/}%
\providecommand \selectlanguage [0]{\@gobble}%
\providecommand \bibinfo  [0]{\@secondoftwo}%
\providecommand \bibfield  [0]{\@secondoftwo}%
\providecommand \translation [1]{[#1]}%
\providecommand \BibitemOpen [0]{}%
\providecommand \bibitemStop [0]{}%
\providecommand \bibitemNoStop [0]{.\EOS\space}%
\providecommand \EOS [0]{\spacefactor3000\relax}%
\providecommand \BibitemShut  [1]{\csname bibitem#1\endcsname}%
\let\auto@bib@innerbib\@empty
\bibitem [{\citenamefont {de~Kemmeter}\ \emph {et~al.}(2025)\citenamefont {de~Kemmeter}, \citenamefont {Ruffo},\ and\ \citenamefont {Gherardini}}]{deKemmeterLetter2025}%
  \BibitemOpen
  \bibfield  {author} {\bibinfo {author} {\bibfnamefont {J.-F.}\ \bibnamefont {de~Kemmeter}}, \bibinfo {author} {\bibfnamefont {S.}~\bibnamefont {Ruffo}},\ and\ \bibinfo {author} {\bibfnamefont {S.}~\bibnamefont {Gherardini}},\ }\href {https://doi.org/10.48550/arXiv.2511.07207} {\bibfield  {journal} {\bibinfo  {journal} {arXiv preprint arXiv:2511.07207}\ } (\bibinfo {year} {2025})}\BibitemShut {NoStop}%
\bibitem [{\citenamefont {Campa}\ \emph {et~al.}(2009)\citenamefont {Campa}, \citenamefont {Dauxois},\ and\ \citenamefont {Ruffo}}]{campa2009statistical}%
  \BibitemOpen
  \bibfield  {author} {\bibinfo {author} {\bibfnamefont {A.}~\bibnamefont {Campa}}, \bibinfo {author} {\bibfnamefont {T.}~\bibnamefont {Dauxois}},\ and\ \bibinfo {author} {\bibfnamefont {S.}~\bibnamefont {Ruffo}},\ }\href {https://doi.org/10.1016/j.physrep.2009.07.001} {\bibfield  {journal} {\bibinfo  {journal} {Phys. Rep.}\ }\textbf {\bibinfo {volume} {480}},\ \bibinfo {pages} {57} (\bibinfo {year} {2009})}\BibitemShut {NoStop}%
\bibitem [{\citenamefont {Defenu}\ \emph {et~al.}(2023)\citenamefont {Defenu}, \citenamefont {Donner}, \citenamefont {Macr\`{\i}}, \citenamefont {Pagano}, \citenamefont {Ruffo},\ and\ \citenamefont {Trombettoni}}]{defenu2023long}%
  \BibitemOpen
  \bibfield  {author} {\bibinfo {author} {\bibfnamefont {N.}~\bibnamefont {Defenu}}, \bibinfo {author} {\bibfnamefont {T.}~\bibnamefont {Donner}}, \bibinfo {author} {\bibfnamefont {T.}~\bibnamefont {Macr\`{\i}}}, \bibinfo {author} {\bibfnamefont {G.}~\bibnamefont {Pagano}}, \bibinfo {author} {\bibfnamefont {S.}~\bibnamefont {Ruffo}},\ and\ \bibinfo {author} {\bibfnamefont {A.}~\bibnamefont {Trombettoni}},\ }\href {https://doi.org/10.1103/RevModPhys.95.035002} {\bibfield  {journal} {\bibinfo  {journal} {Rev. Mod. Phys.}\ }\textbf {\bibinfo {volume} {95}},\ \bibinfo {pages} {035002} (\bibinfo {year} {2023})}\BibitemShut {NoStop}%
\bibitem [{\citenamefont {Dyson}(1969)}]{Dyson1969}%
  \BibitemOpen
  \bibfield  {author} {\bibinfo {author} {\bibfnamefont {F.~J.}\ \bibnamefont {Dyson}},\ }\href {https://doi.org/10.1007/BF01645907} {\bibfield  {journal} {\bibinfo  {journal} {Commun. Math. Phys.}\ }\textbf {\bibinfo {volume} {12}},\ \bibinfo {pages} {91} (\bibinfo {year} {1969})}\BibitemShut {NoStop}%
\bibitem [{\citenamefont {Thouless}(1969)}]{ThoulessPR1969}%
  \BibitemOpen
  \bibfield  {author} {\bibinfo {author} {\bibfnamefont {D.~J.}\ \bibnamefont {Thouless}},\ }\href {https://doi.org/10.1103/PhysRev.187.732} {\bibfield  {journal} {\bibinfo  {journal} {Phys. Rev.}\ }\textbf {\bibinfo {volume} {187}},\ \bibinfo {pages} {732} (\bibinfo {year} {1969})}\BibitemShut {NoStop}%
\bibitem [{\citenamefont {Dyson}(1971)}]{Dyson1971}%
  \BibitemOpen
  \bibfield  {author} {\bibinfo {author} {\bibfnamefont {F.~J.}\ \bibnamefont {Dyson}},\ }\href {https://doi.org/10.1007/BF01645749} {\bibfield  {journal} {\bibinfo  {journal} {Commun. Math. Phys.}\ }\textbf {\bibinfo {volume} {21}},\ \bibinfo {pages} {269} (\bibinfo {year} {1971})}\BibitemShut {NoStop}%
\bibitem [{\citenamefont {Luijten}\ and\ \citenamefont {Bl\"ote}(1997)}]{LuijtenPRB1997}%
  \BibitemOpen
  \bibfield  {author} {\bibinfo {author} {\bibfnamefont {E.}~\bibnamefont {Luijten}}\ and\ \bibinfo {author} {\bibfnamefont {H.~W.~J.}\ \bibnamefont {Bl\"ote}},\ }\href {https://doi.org/10.1103/PhysRevB.56.8945} {\bibfield  {journal} {\bibinfo  {journal} {Phys. Rev. B}\ }\textbf {\bibinfo {volume} {56}},\ \bibinfo {pages} {8945} (\bibinfo {year} {1997})}\BibitemShut {NoStop}%
\bibitem [{\citenamefont {Tomita}(2009)}]{Tomita2009}%
  \BibitemOpen
  \bibfield  {author} {\bibinfo {author} {\bibfnamefont {Y.}~\bibnamefont {Tomita}},\ }\href {https://doi.org/10.1143/JPSJ.78.014002} {\bibfield  {journal} {\bibinfo  {journal} {J. Phys. Soc. Jpn.}\ }\textbf {\bibinfo {volume} {78}},\ \bibinfo {pages} {014002} (\bibinfo {year} {2009})}\BibitemShut {NoStop}%
\bibitem [{\citenamefont {M\"uller}\ and\ \citenamefont {Janke}(2026)}]{muller2025efficient}%
  \BibitemOpen
  \bibfield  {author} {\bibinfo {author} {\bibfnamefont {F.}~\bibnamefont {M\"uller}}\ and\ \bibinfo {author} {\bibfnamefont {W.}~\bibnamefont {Janke}},\ }\href {https://doi.org/10.1103/29yw-9grr} {\bibfield  {journal} {\bibinfo  {journal} {Phys. Rev. Lett.}\ }\textbf {\bibinfo {volume} {136}},\ \bibinfo {pages} {227101} (\bibinfo {year} {2026})}\BibitemShut {NoStop}%
\bibitem [{\citenamefont {Campa}\ \emph {et~al.}(2014)\citenamefont {Campa}, \citenamefont {Dauxois}, \citenamefont {Fanelli},\ and\ \citenamefont {Ruffo}}]{campa2014physics}%
  \BibitemOpen
  \bibfield  {author} {\bibinfo {author} {\bibfnamefont {A.}~\bibnamefont {Campa}}, \bibinfo {author} {\bibfnamefont {T.}~\bibnamefont {Dauxois}}, \bibinfo {author} {\bibfnamefont {D.}~\bibnamefont {Fanelli}},\ and\ \bibinfo {author} {\bibfnamefont {S.}~\bibnamefont {Ruffo}},\ }\href@noop {} {\emph {\bibinfo {title} {Physics of long-range interacting systems}}}\ (\bibinfo  {publisher} {OUP Oxford},\ \bibinfo {year} {2014})\BibitemShut {NoStop}%
\bibitem [{\citenamefont {Nagle}(1970)}]{nagle1970ising}%
  \BibitemOpen
  \bibfield  {author} {\bibinfo {author} {\bibfnamefont {J.~F.}\ \bibnamefont {Nagle}},\ }\href {https://doi.org/10.1103/PhysRevA.2.2124} {\bibfield  {journal} {\bibinfo  {journal} {Phys. Rev. A}\ }\textbf {\bibinfo {volume} {2}},\ \bibinfo {pages} {2124} (\bibinfo {year} {1970})}\BibitemShut {NoStop}%
\bibitem [{\citenamefont {Kardar}(1983)}]{kardar1983crossover}%
  \BibitemOpen
  \bibfield  {author} {\bibinfo {author} {\bibfnamefont {M.}~\bibnamefont {Kardar}},\ }\href {https://doi.org/10.1103/PhysRevB.28.244} {\bibfield  {journal} {\bibinfo  {journal} {Phys. Rev. B}\ }\textbf {\bibinfo {volume} {28}},\ \bibinfo {pages} {244} (\bibinfo {year} {1983})}\BibitemShut {NoStop}%
\bibitem [{\citenamefont {Mukamel}\ \emph {et~al.}(2005)\citenamefont {Mukamel}, \citenamefont {Ruffo},\ and\ \citenamefont {Schreiber}}]{mukamel2005breaking}%
  \BibitemOpen
  \bibfield  {author} {\bibinfo {author} {\bibfnamefont {D.}~\bibnamefont {Mukamel}}, \bibinfo {author} {\bibfnamefont {S.}~\bibnamefont {Ruffo}},\ and\ \bibinfo {author} {\bibfnamefont {N.}~\bibnamefont {Schreiber}},\ }\href {https://doi.org/10.1103/PhysRevLett.95.240604} {\bibfield  {journal} {\bibinfo  {journal} {Phys. Rev. Lett.}\ }\textbf {\bibinfo {volume} {95}},\ \bibinfo {pages} {240604} (\bibinfo {year} {2005})}\BibitemShut {NoStop}%
\bibitem [{\citenamefont {Campa}\ \emph {et~al.}(2025)\citenamefont {Campa}, \citenamefont {Hovhannisyan}, \citenamefont {Ruffo},\ and\ \citenamefont {Trombettoni}}]{CampaJPAMT2025}%
  \BibitemOpen
  \bibfield  {author} {\bibinfo {author} {\bibfnamefont {A.}~\bibnamefont {Campa}}, \bibinfo {author} {\bibfnamefont {V.}~\bibnamefont {Hovhannisyan}}, \bibinfo {author} {\bibfnamefont {S.}~\bibnamefont {Ruffo}},\ and\ \bibinfo {author} {\bibfnamefont {A.}~\bibnamefont {Trombettoni}},\ }\href {https://doi.org/10.1088/1751-8121/ada64c} {\bibfield  {journal} {\bibinfo  {journal} {J. Phys. A: Math. Theor.}\ }\textbf {\bibinfo {volume} {58}},\ \bibinfo {pages} {035005} (\bibinfo {year} {2025})}\BibitemShut {NoStop}%
\bibitem [{\citenamefont {Fisher}\ and\ \citenamefont {Barber}(1972)}]{FisherPRL1972}%
  \BibitemOpen
  \bibfield  {author} {\bibinfo {author} {\bibfnamefont {M.~E.}\ \bibnamefont {Fisher}}\ and\ \bibinfo {author} {\bibfnamefont {M.~N.}\ \bibnamefont {Barber}},\ }\href {https://doi.org/10.1103/PhysRevLett.28.1516} {\bibfield  {journal} {\bibinfo  {journal} {Phys. Rev. Lett.}\ }\textbf {\bibinfo {volume} {28}},\ \bibinfo {pages} {1516} (\bibinfo {year} {1972})}\BibitemShut {NoStop}%
\bibitem [{\citenamefont {Barber}(1983)}]{Barber1983}%
  \BibitemOpen
  \bibfield  {author} {\bibinfo {author} {\bibfnamefont {M.~N.}\ \bibnamefont {Barber}},\ }in\ \href@noop {} {\emph {\bibinfo {booktitle} {Phase Transitions and Critical Phenomena}}},\ Vol.~\bibinfo {volume} {8},\ \bibinfo {editor} {edited by\ \bibinfo {editor} {\bibfnamefont {C.}~\bibnamefont {Domb}}\ and\ \bibinfo {editor} {\bibfnamefont {J.~L.}\ \bibnamefont {Lebowitz}}}\ (\bibinfo  {publisher} {Academic},\ \bibinfo {address} {New York},\ \bibinfo {year} {1983})\ pp.\ \bibinfo {pages} {145--266}\BibitemShut {NoStop}%
\bibitem [{\citenamefont {Privman}(1990)}]{Privman1990Editor}%
  \BibitemOpen
  \bibinfo {editor} {\bibfnamefont {V.}~\bibnamefont {Privman}},\ ed.,\ \href@noop {} {\emph {\bibinfo {title} {{Finite Size Scaling And Numerical Simulation Of Statistical Systems}}}}\ (\bibinfo  {publisher} {World Scientific},\ \bibinfo {address} {Singapore},\ \bibinfo {year} {1990})\BibitemShut {NoStop}%
\bibitem [{\citenamefont {Goldenfeld}(1992)}]{Goldenfeld1992Lectures}%
  \BibitemOpen
  \bibfield  {author} {\bibinfo {author} {\bibfnamefont {N.}~\bibnamefont {Goldenfeld}},\ }\href {https://doi.org/10.1201/9780429493492} {\emph {\bibinfo {title} {{Lectures on Phase Transitions and the Renormalization Group}}}}\ (\bibinfo  {publisher} {CRC Press (Taylor \& Francis Group)},\ \bibinfo {year} {1992})\BibitemShut {NoStop}%
\bibitem [{\citenamefont {Glauber}(1963)}]{glauber1963time}%
  \BibitemOpen
  \bibfield  {author} {\bibinfo {author} {\bibfnamefont {R.~J.}\ \bibnamefont {Glauber}},\ }\href {https://doi.org/10.1063/1.1703954} {\bibfield  {journal} {\bibinfo  {journal} {J. Math. Phys.}\ }\textbf {\bibinfo {volume} {4}},\ \bibinfo {pages} {294} (\bibinfo {year} {1963})}\BibitemShut {NoStop}%
\bibitem [{\citenamefont {Griffiths}\ \emph {et~al.}(1966)\citenamefont {Griffiths}, \citenamefont {Weng},\ and\ \citenamefont {Langer}}]{griffiths1966relaxation}%
  \BibitemOpen
  \bibfield  {author} {\bibinfo {author} {\bibfnamefont {R.~B.}\ \bibnamefont {Griffiths}}, \bibinfo {author} {\bibfnamefont {C.-Y.}\ \bibnamefont {Weng}},\ and\ \bibinfo {author} {\bibfnamefont {J.~S.}\ \bibnamefont {Langer}},\ }\href {https://doi.org/10.1103/PhysRev.149.301} {\bibfield  {journal} {\bibinfo  {journal} {Phys. Rev.}\ }\textbf {\bibinfo {volume} {149}},\ \bibinfo {pages} {301} (\bibinfo {year} {1966})}\BibitemShut {NoStop}%
\bibitem [{\citenamefont {Antoni}\ \emph {et~al.}(2004)\citenamefont {Antoni}, \citenamefont {Ruffo},\ and\ \citenamefont {Torcini}}]{antoni2004first}%
  \BibitemOpen
  \bibfield  {author} {\bibinfo {author} {\bibfnamefont {M.}~\bibnamefont {Antoni}}, \bibinfo {author} {\bibfnamefont {S.}~\bibnamefont {Ruffo}},\ and\ \bibinfo {author} {\bibfnamefont {A.}~\bibnamefont {Torcini}},\ }\href {https://doi.org/10.1209/epl/i2004-10028-6} {\bibfield  {journal} {\bibinfo  {journal} {EPL}\ }\textbf {\bibinfo {volume} {66}},\ \bibinfo {pages} {645} (\bibinfo {year} {2004})}\BibitemShut {NoStop}%
\bibitem [{\citenamefont {Chavanis}(2005)}]{chavanis2005lifetime}%
  \BibitemOpen
  \bibfield  {author} {\bibinfo {author} {\bibfnamefont {P.-H.}\ \bibnamefont {Chavanis}},\ }\href {https://doi.org/10.1051/0004-6361:20041114} {\bibfield  {journal} {\bibinfo  {journal} {Astronomy \& Astrophysics}\ }\textbf {\bibinfo {volume} {432}},\ \bibinfo {pages} {117} (\bibinfo {year} {2005})}\BibitemShut {NoStop}%
\bibitem [{\citenamefont {Chavanis}(2026)}]{chavanis2026thermal}%
  \BibitemOpen
  \bibfield  {author} {\bibinfo {author} {\bibfnamefont {P.-H.}\ \bibnamefont {Chavanis}},\ }\href@noop {} {\bibfield  {journal} {\bibinfo  {journal} {arXiv preprint arXiv:2605.03771}\ } (\bibinfo {year} {2026})}\BibitemShut {NoStop}%
\bibitem [{\citenamefont {Antal}\ \emph {et~al.}(2004)\citenamefont {Antal}, \citenamefont {Droz},\ and\ \citenamefont {R{\'a}cz}}]{antal2004probability}%
  \BibitemOpen
  \bibfield  {author} {\bibinfo {author} {\bibfnamefont {T.}~\bibnamefont {Antal}}, \bibinfo {author} {\bibfnamefont {M.}~\bibnamefont {Droz}},\ and\ \bibinfo {author} {\bibfnamefont {Z.}~\bibnamefont {R{\'a}cz}},\ }\href {https://doi.org/10.1088/0305-4470/37/5/001} {\bibfield  {journal} {\bibinfo  {journal} {J. Phys. A: Math. Gen.}\ }\textbf {\bibinfo {volume} {37}},\ \bibinfo {pages} {1465} (\bibinfo {year} {2004})}\BibitemShut {NoStop}%
\bibitem [{\citenamefont {Abramowitz}\ and\ \citenamefont {Stegun}(1965)}]{abramowitz1965handbook}%
  \BibitemOpen
  \bibfield  {author} {\bibinfo {author} {\bibfnamefont {M.}~\bibnamefont {Abramowitz}}\ and\ \bibinfo {author} {\bibfnamefont {I.~A.}\ \bibnamefont {Stegun}},\ }\href@noop {} {\emph {\bibinfo {title} {Handbook of mathematical functions: with formulas, graphs, and mathematical tables}}},\ Vol.~\bibinfo {volume} {55}\ (\bibinfo  {publisher} {Courier Corporation},\ \bibinfo {year} {1965})\BibitemShut {NoStop}%
\bibitem [{\citenamefont {Touchette}(2009)}]{touchette2009large}%
  \BibitemOpen
  \bibfield  {author} {\bibinfo {author} {\bibfnamefont {H.}~\bibnamefont {Touchette}},\ }\href {https://doi.org/10.1016/j.physrep.2009.05.002} {\bibfield  {journal} {\bibinfo  {journal} {Phys. Rep.}\ }\textbf {\bibinfo {volume} {478}},\ \bibinfo {pages} {1} (\bibinfo {year} {2009})}\BibitemShut {NoStop}%
\bibitem [{\citenamefont {Dantchev}\ \emph {et~al.}(2024)\citenamefont {Dantchev}, \citenamefont {Tonchev},\ and\ \citenamefont {Rudnick}}]{Dantchev2024}%
  \BibitemOpen
  \bibfield  {author} {\bibinfo {author} {\bibfnamefont {D.}~\bibnamefont {Dantchev}}, \bibinfo {author} {\bibfnamefont {N.}~\bibnamefont {Tonchev}},\ and\ \bibinfo {author} {\bibfnamefont {J.}~\bibnamefont {Rudnick}},\ }\href {https://doi.org/10.1103/PhysRevE.110.L062104} {\bibfield  {journal} {\bibinfo  {journal} {Phys. Rev. E}\ }\textbf {\bibinfo {volume} {110}},\ \bibinfo {pages} {L062104} (\bibinfo {year} {2024})}\BibitemShut {NoStop}%
\bibitem [{\citenamefont {Gaspard}(2012)}]{gaspard2012fluctuation}%
  \BibitemOpen
  \bibfield  {author} {\bibinfo {author} {\bibfnamefont {P.}~\bibnamefont {Gaspard}},\ }\href {https://doi.org/10.1088/1742-5468/2012/08/P08021} {\bibfield  {journal} {\bibinfo  {journal} {J. Stat. Mech.}\ }\textbf {\bibinfo {volume} {2012}},\ \bibinfo {pages} {P08021} (\bibinfo {year} {2012})}\BibitemShut {NoStop}%
\bibitem [{\citenamefont {Baxter}(1985)}]{baxter1985exactly}%
  \BibitemOpen
  \bibfield  {author} {\bibinfo {author} {\bibfnamefont {R.~J.}\ \bibnamefont {Baxter}},\ }in\ \href@noop {} {\emph {\bibinfo {booktitle} {Integrable systems in statistical mechanics}}}\ (\bibinfo  {publisher} {World Scientific},\ \bibinfo {year} {1985})\ pp.\ \bibinfo {pages} {5--63}\BibitemShut {NoStop}%
\bibitem [{\citenamefont {Wang}\ \emph {et~al.}(1995)\citenamefont {Wang}, \citenamefont {Hatano},\ and\ \citenamefont {Suzuki}}]{wang1995study}%
  \BibitemOpen
  \bibfield  {author} {\bibinfo {author} {\bibfnamefont {F.}~\bibnamefont {Wang}}, \bibinfo {author} {\bibfnamefont {N.}~\bibnamefont {Hatano}},\ and\ \bibinfo {author} {\bibfnamefont {M.}~\bibnamefont {Suzuki}},\ }\href {https://doi.org/10.1088/0305-4470/28/16/012} {\bibfield  {journal} {\bibinfo  {journal} {J. Phys. A: Math. Gen.}\ }\textbf {\bibinfo {volume} {28}},\ \bibinfo {pages} {4543} (\bibinfo {year} {1995})}\BibitemShut {NoStop}%
\bibitem [{\citenamefont {Ma}(2001)}]{ma2018modern}%
  \BibitemOpen
  \bibfield  {author} {\bibinfo {author} {\bibfnamefont {S.-K.}\ \bibnamefont {Ma}},\ }\href {https://doi.org/10.4324/9780429498886} {\emph {\bibinfo {title} {{Modern Theory Of Critical Phenomena}}}}\ (\bibinfo  {publisher} {Routledge},\ \bibinfo {year} {2001})\BibitemShut {NoStop}%
\bibitem [{\citenamefont {Ellis}\ and\ \citenamefont {Newman}(1978)}]{Ellis1978}%
  \BibitemOpen
  \bibfield  {author} {\bibinfo {author} {\bibfnamefont {R.~S.}\ \bibnamefont {Ellis}}\ and\ \bibinfo {author} {\bibfnamefont {C.~M.}\ \bibnamefont {Newman}},\ }\href {https://doi.org/10.1007/BF00533049} {\bibfield  {journal} {\bibinfo  {journal} {Z. Wahrscheinlichkeitstheorie verw Gebiete}\ }\textbf {\bibinfo {volume} {44}},\ \bibinfo {pages} {117} (\bibinfo {year} {1978})}\BibitemShut {NoStop}%
\bibitem [{\citenamefont {Ellis}\ \emph {et~al.}(1980)\citenamefont {Ellis}, \citenamefont {Newman},\ and\ \citenamefont {Rosen}}]{Ellis1980}%
  \BibitemOpen
  \bibfield  {author} {\bibinfo {author} {\bibfnamefont {R.~S.}\ \bibnamefont {Ellis}}, \bibinfo {author} {\bibfnamefont {C.~M.}\ \bibnamefont {Newman}},\ and\ \bibinfo {author} {\bibfnamefont {J.~S.}\ \bibnamefont {Rosen}},\ }\href {https://doi.org/10.1007/BF00536186} {\bibfield  {journal} {\bibinfo  {journal} {Z. Wahrscheinlichkeitstheorie verw Gebiete}\ }\textbf {\bibinfo {volume} {51}},\ \bibinfo {pages} {153} (\bibinfo {year} {1980})}\BibitemShut {NoStop}%
\bibitem [{\citenamefont {Kenna}(2004)}]{Kenna2004}%
  \BibitemOpen
  \bibfield  {author} {\bibinfo {author} {\bibfnamefont {R.}~\bibnamefont {Kenna}},\ }\href {https://doi.org/https://doi.org/10.1016/j.nuclphysb.2004.05.012} {\bibfield  {journal} {\bibinfo  {journal} {Nucl. Phys. B}\ }\textbf {\bibinfo {volume} {691}},\ \bibinfo {pages} {292} (\bibinfo {year} {2004})}\BibitemShut {NoStop}%
\bibitem [{\citenamefont {Honchar}\ \emph {et~al.}(2024)\citenamefont {Honchar}, \citenamefont {Berche}, \citenamefont {Holovatch},\ and\ \citenamefont {Kenna}}]{Honchar2024}%
  \BibitemOpen
  \bibfield  {author} {\bibinfo {author} {\bibfnamefont {Y.}~\bibnamefont {Honchar}}, \bibinfo {author} {\bibfnamefont {B.}~\bibnamefont {Berche}}, \bibinfo {author} {\bibfnamefont {Y.}~\bibnamefont {Holovatch}},\ and\ \bibinfo {author} {\bibfnamefont {R.}~\bibnamefont {Kenna}},\ }\href {https://doi.org/10.5488/CMP.27.13603} {\bibfield  {journal} {\bibinfo  {journal} {Condens. Matter Phys.}\ }\textbf {\bibinfo {volume} {27}},\ \bibinfo {pages} {13603} (\bibinfo {year} {2024})}\BibitemShut {NoStop}%
\bibitem [{\citenamefont {Hohenberg}\ and\ \citenamefont {Halperin}(1977)}]{hohenberg1977theory}%
  \BibitemOpen
  \bibfield  {author} {\bibinfo {author} {\bibfnamefont {P.~C.}\ \bibnamefont {Hohenberg}}\ and\ \bibinfo {author} {\bibfnamefont {B.~I.}\ \bibnamefont {Halperin}},\ }\href {https://doi.org/10.1103/RevModPhys.49.435} {\bibfield  {journal} {\bibinfo  {journal} {Rev. Mod. Phys.}\ }\textbf {\bibinfo {volume} {49}},\ \bibinfo {pages} {435} (\bibinfo {year} {1977})}\BibitemShut {NoStop}%
\bibitem [{\citenamefont {Risken}(1989)}]{risken1989fokker}%
  \BibitemOpen
  \bibfield  {author} {\bibinfo {author} {\bibfnamefont {H.}~\bibnamefont {Risken}},\ }in\ \href@noop {} {\emph {\bibinfo {booktitle} {{The Fokker-Planck equation: methods of solution and applications}}}}\ (\bibinfo  {publisher} {Springer},\ \bibinfo {year} {1989})\ pp.\ \bibinfo {pages} {63--95}\BibitemShut {NoStop}%
\bibitem [{\citenamefont {Mori}\ \emph {et~al.}(2010)\citenamefont {Mori}, \citenamefont {Miyashita},\ and\ \citenamefont {Rikvold}}]{mori2010asymptotic}%
  \BibitemOpen
  \bibfield  {author} {\bibinfo {author} {\bibfnamefont {T.}~\bibnamefont {Mori}}, \bibinfo {author} {\bibfnamefont {S.}~\bibnamefont {Miyashita}},\ and\ \bibinfo {author} {\bibfnamefont {P.~A.}\ \bibnamefont {Rikvold}},\ }\href {https://doi.org/10.1103/PhysRevE.81.011135} {\bibfield  {journal} {\bibinfo  {journal} {Phys. Rev. E}\ }\textbf {\bibinfo {volume} {81}},\ \bibinfo {pages} {011135} (\bibinfo {year} {2010})}\BibitemShut {NoStop}%
\bibitem [{\citenamefont {Saadat}\ \emph {et~al.}(2023)\citenamefont {Saadat}, \citenamefont {Latella},\ and\ \citenamefont {Ruffo}}]{Saadat2023}%
  \BibitemOpen
  \bibfield  {author} {\bibinfo {author} {\bibfnamefont {E.}~\bibnamefont {Saadat}}, \bibinfo {author} {\bibfnamefont {I.}~\bibnamefont {Latella}},\ and\ \bibinfo {author} {\bibfnamefont {S.}~\bibnamefont {Ruffo}},\ }\href {https://doi.org/10.1088/1742-5468/acecf9} {\bibfield  {journal} {\bibinfo  {journal} {J. Stat. Mech.: Theory Exp.}\ }\textbf {\bibinfo {volume} {2023}}\bibinfo  {number} { (8)},\ \bibinfo {pages} {083207}}\BibitemShut {NoStop}%
\bibitem [{\citenamefont {Lee}\ and\ \citenamefont {Seo}(2022)}]{lee2022non}%
  \BibitemOpen
\bibfield  {number} {  }\bibfield  {author} {\bibinfo {author} {\bibfnamefont {J.}~\bibnamefont {Lee}}\ and\ \bibinfo {author} {\bibfnamefont {I.}~\bibnamefont {Seo}},\ }\href {https://doi.org/10.1007/s00440-021-01102-z} {\bibfield  {journal} {\bibinfo  {journal} {Probab. Theory Relat. Fields}\ }\textbf {\bibinfo {volume} {182}},\ \bibinfo {pages} {849} (\bibinfo {year} {2022})}\BibitemShut {NoStop}%
\bibitem [{\citenamefont {Cvitkovi{\'c}}\ \emph {et~al.}(2017)\citenamefont {Cvitkovi{\'c}}, \citenamefont {Smith},\ and\ \citenamefont {Pande}}]{cvitkovic2017asymptotic}%
  \BibitemOpen
  \bibfield  {author} {\bibinfo {author} {\bibfnamefont {M.}~\bibnamefont {Cvitkovi{\'c}}}, \bibinfo {author} {\bibfnamefont {A.-S.}\ \bibnamefont {Smith}},\ and\ \bibinfo {author} {\bibfnamefont {J.}~\bibnamefont {Pande}},\ }\href {https://doi.org/10.1088/1751-8121/aa7213} {\bibfield  {journal} {\bibinfo  {journal} {J. Phys. A: Math. Theor.}\ }\textbf {\bibinfo {volume} {50}},\ \bibinfo {pages} {265206} (\bibinfo {year} {2017})}\BibitemShut {NoStop}%
\end{thebibliography}%

\end{document}